\documentclass[aps,pra,twocolumn,footinbib,longbibliography]{revtex4-2}
\usepackage{graphicx}
\usepackage[utf8]{inputenc}
\usepackage{indentfirst}
\usepackage{physics}
\usepackage{braket}
\usepackage{float}
\usepackage{amsmath}
\usepackage{epstopdf}
\usepackage{footnote} 
\usepackage{CJK}
\usepackage{esint}
\usepackage{color}
\usepackage[T1]{fontenc}
\usepackage{subfigure}
\usepackage{amsfonts}
\usepackage{footmisc}
\usepackage{scrextend}
\usepackage{multirow}
\usepackage[hyperfootnotes=false]{hyperref}
\usepackage[acronym]{glossaries}

\usepackage[english]{babel}
\usepackage{url}
\usepackage{bm}
\usepackage{hyperref}
\definecolor{darkblue}{rgb}{0,0,0.5}
\hypersetup{
colorlinks=true,
linkcolor=black,
filecolor=blue,
citecolor=darkblue,  
urlcolor=black,
}

\newcommand{\bolda}{\boldsymbol {a}}
\newcommand{\bx}{\boldsymbol x}
\newcommand{\bxi}{\boldsymbol \xi}
\newcommand{\balpha}{\boldsymbol \alpha}
\newcommand{\bOmega}{\boldsymbol \Omega}
\newcommand{\bI}{\boldsymbol I}
\newcommand{\bZ}{\boldsymbol Z}

\newcommand*\diff{\mathop{}\!\mathrm{d}}

\newcommand {\abold}{\ensuremath \hat{\boldsymbol{a}}}

\newcommand{\calA}{{\cal A}}

\newcommand{\calL}{{\cal L}}
\newcommand{\calN}{{\cal N}} 
\newcommand{\calT}{{\cal T}}

\newcommand{\calH}{{\cal H}}

\newcommand{\1}{^{(1)}}

\def\be{\begin{equation}}
\def\ee{\end{equation}}
\def\ba{\begin{eqnarray}}
\def\ea{\end{eqnarray}}
\usepackage{bm}

\newcommand{\QZ}[1]{{{\textcolor{black}{#1}}}}
\newcommand{\JW}[1]{{{\textcolor{black}{#1}}}}

\begin{document}

\title{Deterministic microwave-optical transduction based on quantum teleportation}
\author{Jing Wu$^{1}$}
\author{Chaohan Cui$^{1}$}
\author{Linran Fan$^{1}$}
\author{Quntao Zhuang$^{1,2}$}
\email{zhuangquntao@email.arizona.edu}

\address{
$^1$James C. Wyant College of Optical Sciences
}
\address{
$^2$Department of Electrical and Computer Engineering, University of Arizona, Tucson, Arizona 85721, USA
}

\begin{abstract}
The coherent transduction between microwave and optical frequencies is critical to interconnect superconducting quantum processors over long distances. However, it is challenging to establish such a quantum interface with high efficiency and small added noise based on the standard direct conversion scheme. Here, we propose an electro-optic transduction system based on continuous-variable quantum teleportation. Reliable quantum information transmission can be realized with an arbitrarily small cooperativity, in contrast to the direct conversion scheme which requires a large minimum cooperativity. We show that the teleportation-based scheme maintains a significant rate advantage robustly for practical thermal noise and all values of cooperativity. We further investigate the performance in the transduction of complex quantum states such as cat states and Gottesman-Kitaev-Preskill(GKP) states and show that a higher fidelity or success probability can be achieved with the teleportation-based scheme. Our scheme significantly reduces the device requirement, and makes quantum transduction between microwave and optical frequencies feasible in the near future.
\end{abstract}

\maketitle

Quantum networks~\cite{kimble2008quantum,biamonte2019complex,wehner2018quantum,kozlowski2019towards,zhang2021entanglement} have been envisioned as high-performance quantum processors interconnected by efficient quantum communication channels. Quantum processors require strong nonlinear interaction at single-quanta level, which can be readily realized with Josephson effect at microwave frequencies in superconducting circuits~\cite{campagne2020quantum,wang2020efficient,elder2020high,heeres2017implementing}. 
However, the high attenuation and thermal noise at room temperature prevent the direct transmission of quantum states at microwave frequencies over long distances. In contrast, optical photons are the ideal candidate to transmit quantum information over long distances with the near-zero thermal noise and low attenuation at room temperature~\cite{Bennett20147,Ekert_1991,gisin2002,xu2020,pirandola2020advances}. However, it is challenging to develop high-fidelity deterministic quantum gates due to the weak optical nonlinearity. The complementary characteristics of microwave and optical photons calls for a hybrid quantum platform where quantum information is processed by superconducting circuits and transmitted with optical photons. Therefore, an efficient scheme to interconvert quantum states between microwave and optical photons is of paramount importance~\cite{lauk2020perspectives,awschalom2021development,vainsencher2016bi,balram2016coherent,fan2018superconducting,shao2019microwave,han2020cavity,zhong2020proposal,mirhosseini2020superconducting,forsch2020microwave,jiang2020efficient,fiaschi2021optomechanical}.

The coherent conversion of quantum states between microwave and optical frequencies have been proposed using various platforms including nanomechanics~\cite{andrews2014bidirectional,bochmann2013nanomechanical}, electro-optics~\cite{Tsang2010,Tsang2011,fan2018superconducting,xu2020bidirectional}, magnons~\cite{hisatomi2016bidirectional,williamson2014magneto}, rare-earth-ion crystals~\cite{bartholomew2020chip}, and cold atoms~\cite{hafezi2012atomic}. Regardless of the physical implementations, an interaction Hamiltonian performing beam-splitter operations in the frequency domain is used in all schemes to directly transduce quantum states between microwave and optical frequencies. However, it is still challenging to realize a transduction system with high efficiency and low added noise, which are required for high-fidelity quantum state conversion. Other than direct transmission, quantum communications can also be realized efficiently with shared entanglement and classical communication~\cite{bennett1993,braunstein1998,pirandola2006,barzanjeh2012reversible,zhang2018quantum,lau2019high,rueda2019electro,zhong2020proposal}. A recent study~\cite{zhong2020proposal} proposed a scheme to establish time-bin entanglement and perform teleportation to transfer time-bin encoded qubits between microwave and optical frequencies. However, unlike direct conversion, the simple time-bin entanglement is incapable of transducing complex quantum states such as cat states and Gottesman–Kitaev-Preskill (GKP) states~\cite{gottesman2001}, which are important for robust quantum operations against loss and noise~\cite{gottesman2001,noh2019encoding,fluhmann2019encoding,campagne2020quantum,grimm2020stabilization,rozpkedek2020quantum,fukui2020all,ma2021quantum,wu2021}. The probabilistic nature of time-bin entanglement generation also renders difficulty in achieving a high conversion rate. This issue can be resolved by using continuous-variable quantum teleportation~\cite{braunstein1998,pirandola2006,barzanjeh2012reversible,rueda2019electro}.

In this paper, we propose an electro-optic transduction system to enable the conversion of complex quantum states between microwave and optical frequencies based on continuous-variable quantum teleportation. To begin with, we show that direct conversion completely fails such a task at small cooperativity due to zero quantum capacity~\cite{quantum_capacity_Lloyd,quantum_capacity_Shor,quantum_capacity_Devetak}. On the contrary, our teleportation-based scheme demonstrates appreciable rates of quantum state conversion under the same condition and provides a strict rate advantage in a wide range of device parameters. In particular, our system can work at an operating temperature of \QZ{0.2 K}, with robust performance against thermal noise. This is achieved by the elimination of intermediate excitations, in contrast to other systems. We further consider the transduction performance of three states that are widely used in quantum information processing, the coherent state, cat state and finite-squeezed GKP state~\cite{gottesman2001}, where large advantages can be found under practical experimental conditions. All required operations in the teleportation-based transduction scheme can be readily realized in both microwave and optical domains, making efficient microwave-optical transduction possible with current experimental conditions.

\section{Cavity electro-optics} 
\begin{figure}
    \centering
    \includegraphics[width=0.45\textwidth]{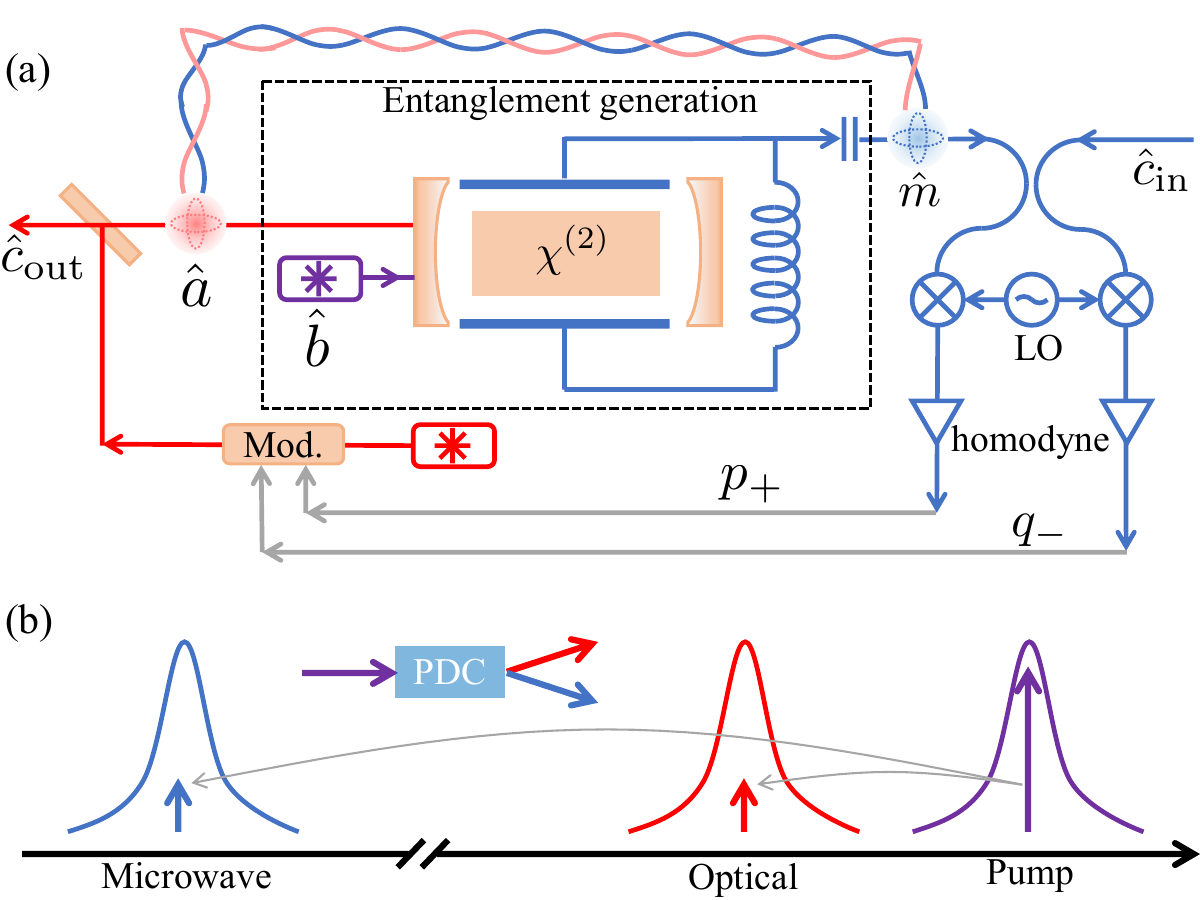}
    \caption{(a) Schematic of the teleportation based transduction scheme. The dashed box indicates the entanglement generation between the microwave mode $\hat{m}$ and the optical mode $\hat{a}$, in a three-wave mixing described by Hamiltonian in Eq.~\eqref{H_all} jointly with optical pump mode $\hat{b}$. (b) Entanglement generation between microwave and optical domains. Purple indicates the optical pump, blue indicates the microwave modes and red indicates optical modes. LO: local oscillator. Mod.: modulator. PDC: parametric down conversion. 
    }
    \label{fig:schematic}
\end{figure}

Superconducting cavity electro-optics~\cite{Tsang2010,Tsang2011,fan2018superconducting,xu2020bidirectional,fu2021cavity} is one of the most promising platforms for on-chip microwave-optical transduction, as it directly converts quantum states and eliminates noisy intermediate excitations in other platforms.
%Different from other microwave-optical transduction platforms that rely on intermediate excitations, cavity electro-optics provide the unique capability to directly convert quantum states between microwave and optical frequencies.
%Therefore, the noise introduced by intermediate excitations can be eliminated. By integrating superconducting microwave resonators with photonic ring cavities, significant improvement over efficiency and added noise has been achieved~\cite{fan2018superconducting,xu2020bidirectional,fu2021cavity}, making superconducting cavity electro-optics one of the most promising platforms for on-chip microwave-optical transduction. 
Such a system can be realized with the setup shown in Fig.~\ref{fig:schematic}(a). The optical cavity consists of materials with $\chi^{(2)}$ nonlinearity, and is placed between the capacitors of a LC microwave resonator. The electric field across the capacitor changes the refractive index of the optical cavity, thus modulating the optical resonant frequency. Reversely, modulated optical fields can generate microwave field due to the optical mixing (rectification) in $\chi^{(2)}$ material. The interacton Hamiltonian of cavity electro-optics has the standard three-wave mixing form
\begin{align}
    H = i\hbar (g\hat{a}^{\dagger}\hat{b}\hat{m}^{\dagger}-g^*\hat{a}\hat{b}^{\dagger}\hat{m}),
    \label{H_all}
\end{align}
with two optical modes ($\hat{a}$ and $\hat{b}$) and one microwave mode ($\hat{m}$). Here $g$ is the coupling coefficient and $\hbar$ is the Planck's constant. If the optical mode $\hat{a}$ is coherently pumped, a beamsplitter interaction Hamiltonian can be realized between the optical mode $\hat{b}$ and the microwave mode $\hat{m}$ for direction conversion. If the optical mode $\hat{b}$ is coherently pumped, a two-mode-squeezing interaction Hamiltonian can be realized between the optical mode $\hat{a}$ and the microwave mode $\hat{m}$ for entanglement generation (see Appendix~\ref{app:generation}). %Sec. II of Ref.~\cite{supp} 
The optical (microwave) modes have intrinsic, coupling, and total loss rates $\gamma_{\rm oi}$, $\gamma_{\rm oc}$, and $\gamma_{\rm o}=\gamma_{\rm oi}+\gamma_{\rm oc}$ ($\gamma_{\rm mi}$, $\gamma_{\rm mc}$, and $\gamma_{\rm m}=\gamma_{\rm mi}+\gamma_{\rm mc}$) respectively. The extraction efficiency for the optical (microwave) mode is defined as $\zeta_{\rm o}=\gamma_{\rm oc}/\gamma_{\rm o}$ ($\zeta_{\rm m}=\gamma_{\rm mc}/\gamma_{\rm m}$). The interaction cooperativity is defined as 
$
C=4 N g^2/{\gamma_{\rm o} \gamma_{\rm m}}
$,
with $N$ the total intra-cavity pump photon number. The stable operation of entanglement generation requires ${C}\in [0,1)$~\cite{Tsang2011}.

The optical thermal noise is neglected in our analysis due to its small occupation even at room temperature. The microwave thermal noise has non-zero mean occupation number $n_{\rm in}$ following the Bose-Einstein distribution.
%The transduction from microwave to optical frequencies is calculated here, and the reverse direction will be similar. 
%\LF{this should be moved to the place when describing Fig. 2?}

\section{Direct conversion} 
In the direct conversion (DC) approach, the transduction device implements an interaction Hamiltonian with a beam-splitter form. Therefore, the direct conversion process can be modeled as a bosonic thermal-attenuator described by the input-output relation
\be 
\QZ{\hat{b}=\sqrt{\eta_{\rm DC}}\hat{m}+\sqrt{1-\eta_{\rm DC}}\hat{e}},
\label{a_DC}
\ee 
with the overall transduction efficiency~\cite{Tsang2011}
\begin{align}
\eta_{\rm DC}=\zeta_{\rm m} \zeta_{\rm o} \frac{4{{C}}}{(1+{{C}})^2},
\label{eta_DC}
\end{align}
and $\hat{e}$ is a thermal mode with mean photon number
\begin{align}
N_{\rm DC}=\frac{1}{1-\eta_{\rm DC}}(1-\zeta_{\rm m}) \zeta_{\rm o} \frac{4{{C}}}{(1+{{C}})^2} n_{\rm in}.
\label{N_DC}
\end{align}
The quantum channel described by Eq.~\eqref{a_DC} can have a nonzero quantum capacity only when $\eta_{\rm DC}> 1/2$~\cite{caruso2006degradability}, which places a threshold for the cooperativity,
\be 
{{C}}\ge -1+4\zeta_{\rm m} \zeta_{\rm o}-\sqrt{8\zeta_{\rm m} \zeta_{\rm o} (2 \zeta_{\rm m} \zeta_{\rm o}-1)}\ge 3-2\sqrt{2}.
\label{C_DC_requirement}
\ee
Even the minimum value is beyond the state-of-the-art experimental demonstrations~\cite{fan2018superconducting,xu2020bidirectional}, therefore forbids the reliable transduction of quantum states. We emphasize that this requirement for direct conversion is essential and cannot be circumvented by error-correction efforts~\cite{noh2019encoding,wu2021}.

\section{Transduction with quantum teleportation} To go beyond direct conversion, we propose to realize the transduction between microwave and optical frequencies with continuous-variable quantum teleportation. In this case, the transduction device is used for two-mode-squeezing generation between the optical and microwave modes (Fig.~\ref{fig:schematic}). The intuition behind this teleportation-based transduction is that classical communication can boost the quantum information transmission rate beyond the unassisted quantum capacity~\cite{bennett1997capacities,pirandola2017fundamental}.

For the transduction from microwave to optical frequencies, the microwave input mode $\hat{c}_{\rm in}$ and the microwave mode $\hat{m}$ of the entangled state are interfered on a 50/50 beamsplitter. The beamsplitter outputs are measured along conjugate quadratures with homodyne detection. The measurement results are used to perform displacement operation on the optical mode $\hat{a}$ of the entangled state with a scaling factor $\kappa$. Then the output optical mode $\hat{c}_{\rm out}$ will be in a state close to the input microwave state $\hat{c}_{\rm in}$ (see Appendices~\ref{app:teleportation} and ~\ref{app:channel_definition}). % Sec. III and Sec. IV of Ref.~\cite{supp}. 
The scaling factor $\kappa$ can be optimized to achieve the best performance. When $\kappa< 1$, the overall channel is a thermal-attenuator described by the input-output relation
\be 
\hat{c}_{\rm out}=\kappa \hat{c}_{\rm in}+\sqrt{1-\kappa^2}\hat{e},
\ee 
When $\kappa>1$, it is a thermal-amplifier:
\be 
\hat{c}_{\rm out}=\kappa \hat{c}_{\rm in}+\sqrt{\kappa^2-1}\hat{e}^\dagger,
\ee 
The thermal mode $\hat{e}$ has the mean photon number
\begin{align}
N_{\rm TP} =\frac{u \kappa^2-2v\kappa + w}{2|1-\kappa^2|}-\frac{1}{2},
\label{N_TP}
\end{align}
with the constants
%where the constants are functions of the cooperativity $C$, the extraction efficiencies $\zeta_{\rm m},\zeta_{\rm o}$ and the thermal noise $n_{\rm in}$:
\JW{
\begin{subequations}
\begin{align}
&u=1+\frac{8\zeta_{\rm m} [{{C}}+ n_{\rm in}(1-\zeta_{\rm m})]}{(1-{{C}})^2},
\\
&v=\frac{4\sqrt{\zeta_{\rm o}\zeta_{\rm m} {{C}}}[1+{{C}}+2 n_{\rm in}(1-\zeta_{\rm m})]}{(1-{{C}})^2},
\\
&w=1+\frac{8{{C}}\zeta_{\rm o}\left[1+n_{\rm in}\left(1-\zeta_{\rm m}\right)\right]}{(1-{{C}})^2}.
\end{align}
\label{uvw}
\end{subequations}
}
When $\kappa = 1$, the channel reduces to an additive white Gaussian noise channel with a variance $(u+w-2v)/2$.

\section{Quantum capacity} 
\begin{figure}
    \centering
    \includegraphics[width=0.45\textwidth]{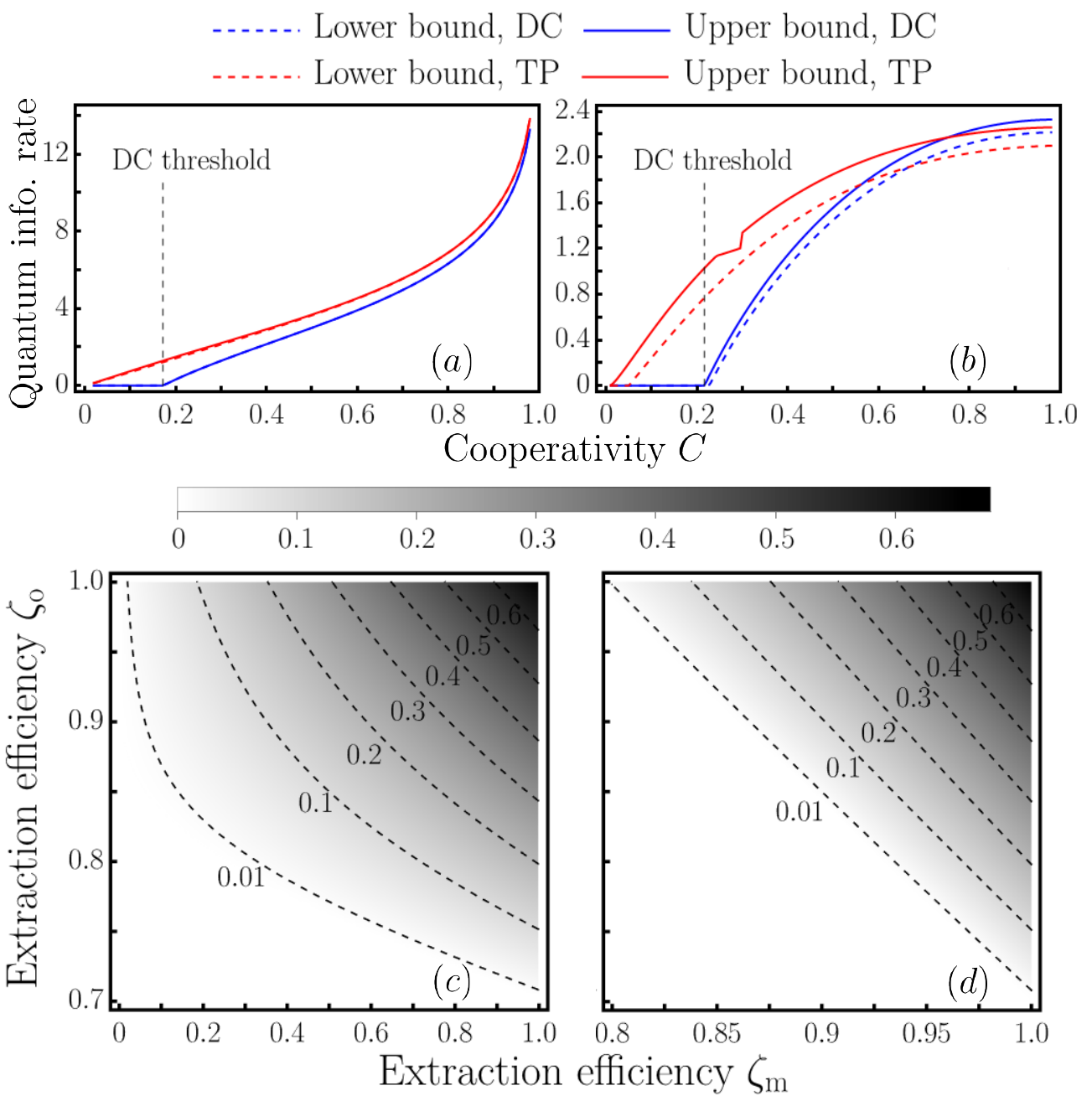}
\caption{Quantum capacity comparison. (a)(b) Capacity bounds versus cooperativity ${{C}}$ with extraction efficiencies (a) $\zeta_{\rm o}=1$, $\zeta_{\rm m}=1$ and (b) $\zeta_{\rm o}=0.9, \zeta_{\rm m}=0.95$. The noise is irrelevant in (a) and \JW{$n_{\rm in} = 0.2$} in (b). The thresholds of the direct conversion in Eq.~\eqref{C_DC_requirement} are indicated by the vertical black dashed lines, with values around (a) $3-2\sqrt{2}\simeq 0.172$ and (b) $0.216$. \JW{We combined multiple different upper bounds (see Appendix \ref{app:q_bound}).}  (c)(d) Contours of the capacity lower bound for the teleportation-based (TP) transduction scheme, with ${{C}}=0.1$ and (c) $n_{\rm in}=0$ (d) \QZ{$n_{\rm in}=0.2$}. In these two cases, the capacity of the direct conversion (DC) scheme is all zero.
\label{fig:capacity}
}
\end{figure}

We first compare the quantum capacities of the two schemes, which provide the ultimate bound of quantum information rates. We begin with the ideal case with ideal extraction efficiencies ($\zeta_{\rm o} = \zeta_{\rm m} = 1$ ) at zero temperature. Therefore, the thermal noise at microwave frequency vanishes and the quantum capacity of direct conversion can be calculated exactly~\cite{wolf2007quantum}. However, the teleportation-based scheme can still have non-zero noise due to finite two-mode squeezing at $C<1$. As the exact solution to quantum capacities for thermal-attenator and thermal-amplifier with non-zero noise is unknown, we calculate the lower bounds~\cite{holevo2001evaluating} and upper bounds~\cite{pirandola2017fundamental,fanizza2021estimating} of quantum capacities instead (Appendix~\ref{app:q_bound}).%(see Sec. V of Ref.~\cite{supp}). 
As shown in Fig.~\ref{fig:capacity}(a), the upper (solid) and lower bound (dashed) coincide exactly for the direct conversion (blue), while a small gap persists for the teleportation scheme (red, see Fig.~\ref{fig:OptK-Qdiff}).%~\cite{supp}. 
The teleportation scheme has a quantum capacity strictly higher than the direct conversion regardless of cooperativity value ${{C}}$. Especially, the quantum capacity for direction conversion is zero when cooperativity is below the threshold (Eq.~\eqref{C_DC_requirement}). In contrast, the teleportation scheme shows non-zero quantum capacity with an arbitrarily small cooperativity. In Fig.~\ref{fig:capacity}(c), we show the quantum capacity lower bound of the teleportation scheme with cooperativity $C=0.1$, which has been experimentally demonstrated~\cite{fan2018superconducting,xu2020bidirectional}. With non-ideal extraction efficiencies, the quantum capacity only decreases gradually, showing the robustness of the teleportation scheme.

Next we consider the practical case with non-ideal extraction efficiencies at finite temperature. In this case, the thermal noise at microwave frequency needs to be considered. We assume the microwave resonator frequency 8~GHz and ambient temperature \QZ{0.2~K}. Therefore, the thermal noise occupation is \QZ{$n_{\rm in}=0.2$}. 
We use $\zeta_{\rm m}=0.95$ and $\zeta_{\rm o}=0.9$, which are experimentally feasible.
The quantum capacity lower bound of the teleportation scheme still has non-zero values regardless of the cooperativity $C$. In contrast, the quantum capacity upper bound of direction conversion is zero when cooperativity is below the threshold (Eq.~\eqref{C_DC_requirement}). Especially, we find that the lower bound of the teleportation scheme is higher than the upper bound of the direct conversion in the low cooperativitiy region, meaning the teleportation scheme strictly outperforms the direction conversion. We also show the quantum capacity of the teleportation scheme with different extraction efficiencies using the condition $C=0.1$ and \JW{$n_{\rm in}=0.2$.} Again, direct transmission has zero quantum capacity in this case. Compared with the ideal $n_{\rm in}=0$ case (Fig.~\ref{fig:capacity}(c)), the quantum capacity drops due to the thermal noise contamination (Fig.~\ref{fig:capacity}(d)). Therefore, a higher microwave extraction efficiency is needed to achieve the same quantum capacity.

\begin{figure}[t]
    \centering
    \includegraphics[width=0.45\textwidth]{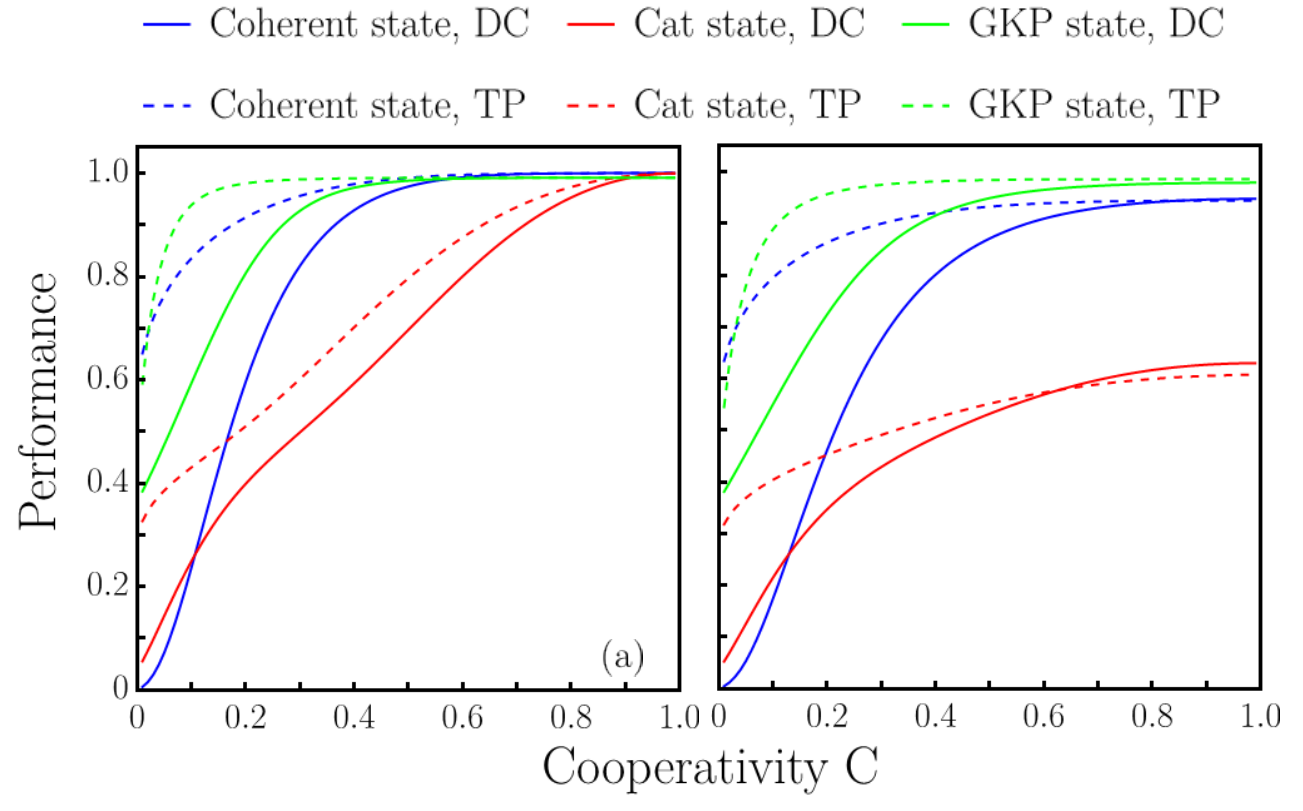}
   \caption{Fidelity for coherent state $\ket{\alpha}$, cat state $N_{+}(\ket{\alpha}+\ket{-\alpha})$ (with $\alpha=2$) and success probability of GKP state transduction. The GKP state has a finite squeezing noise $\sigma_{\rm GKP}=0.22$ (10dB). (a)$\zeta_{\rm m}=1$ and $\zeta_{\rm o}=1$, $n_{\rm in}$ is irrelevant. (b) $\zeta_{\rm m}=0.95$, $\zeta_{\rm o}=0.9$ and \JW{$n_{\rm in}=0.2$.} 
   \label{fig:fidelity_performance}
   }
\end{figure}
\begin{figure}[t]
    \centering
    \includegraphics[width=0.45\textwidth]{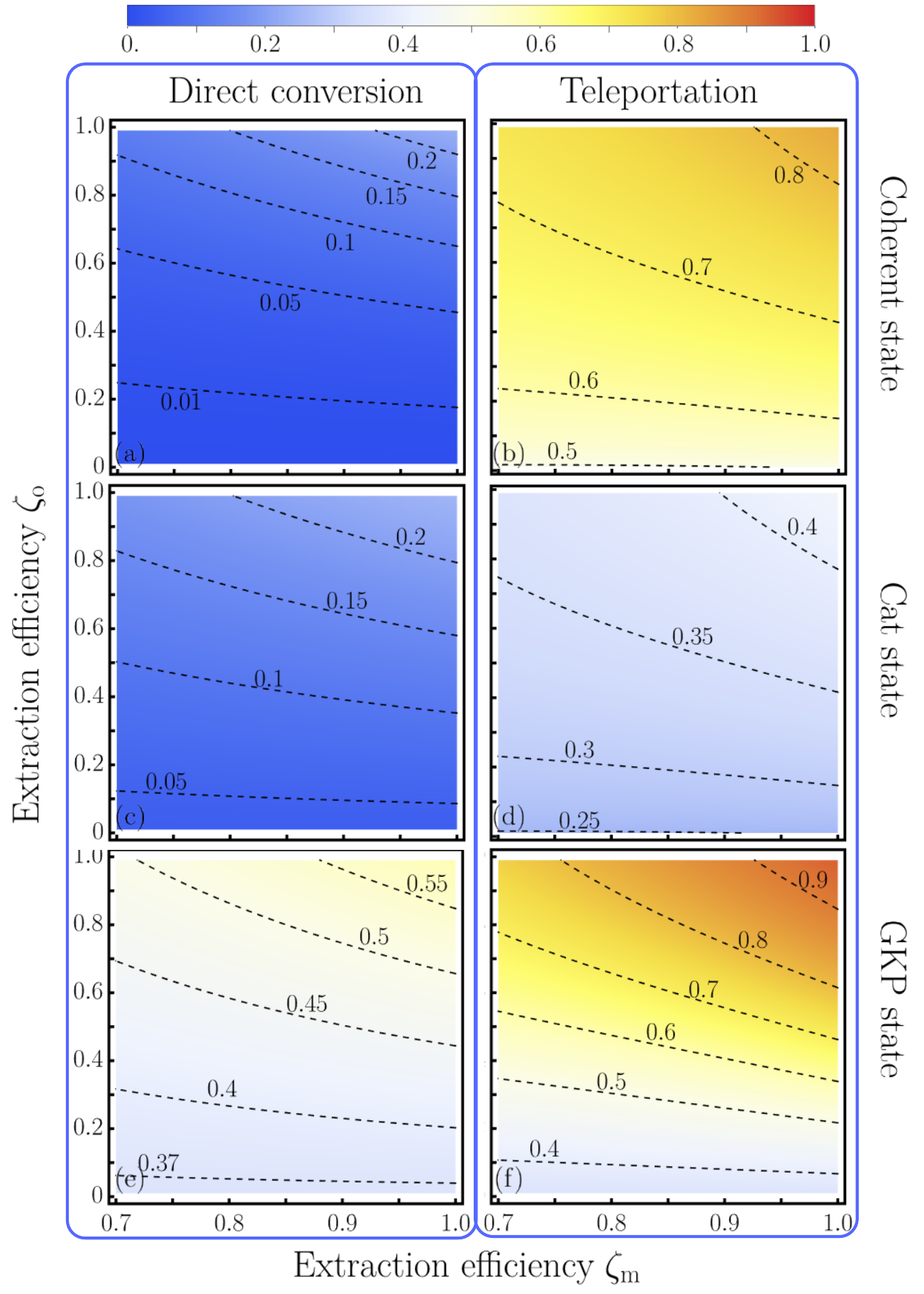}
   \caption{ Performance of quantum state transfer: (a)(b) fidelity for coherent state $\ket{\alpha=2}$, (c)(d) fidelity for cat state $\ket{cat_+}$ with $\alpha=2$ and
   (e)(f) success probability for GKP state with error correction. The GKP state has a finite squeezing noise $\sigma_{\rm GKP}=0.22$ (10 dB). The left column is for direct conversion, while the right column is for teleportation based transduction. We choose ${{C}}=0.1$ and \JW{$n_{\rm in}=0.2$} to represent state-of-the-art experimental condition.
   \label{fig:state}
   }
\end{figure}

\section{Quantum state transfer} 
While quantum capacity shows the ultimate quantum information rate, the transduction performance can vary depending on different quantum states. Here, we consider the quantum states that are important for quantum communication and computation, including the coherent state $\ket{\alpha}$, the cat state $\ket{cat_+}\propto(\ket{\alpha}+\ket{-\alpha})$, and the finite-squeezed GKP states~\cite{gottesman2001,terhal2020towards},
\begin{align} 
\ket{\tilde{k}}_{\rm GKP} \propto \int \diff^2{\alpha} \exp[-\frac{\abs{\alpha}^2}{2\sigma_{\rm{GKP}}^2}]\hat{D}(\alpha)\ket{k}_{\rm GKP}, k=0,1,
\end{align}
where $\hat{D}(\alpha)$ is the displacement operator and $\ket{k}_{\rm GKP}=\sum_{n=-\infty}^\infty \ket{\sqrt{\pi}(2n+k)}_{\hat{q}}$ is the perfect GKP state. The variance $\sigma_{\rm GKP}^2$ characterizes the GKP noise due to finite-squeezing (see Appendix~\ref{app:GKP}). 
%.(see Sec. VIII of Ref.~\cite{supp})

%\equiv \exp(\alpha \hat{a}^\dagger-\alpha^* \hat{a})

For the coherent state and cat state, we directly compare the Uhlmann fidelity $
\mathcal{F}\left(\rho,\sigma\right)=\left({\rm tr}\sqrt{\sqrt{\rho}\sigma\sqrt{\rho}}\right)^2
$~\cite{uhlmann1976transition,jozsa1994fidelity} between the input and the output states (Appendix~\ref{app:fidelity}). %(see Sec. VI of Ref.~\cite{supp}). %. 
In the teleportation-based transduction scheme, the fidelity will depend on the homodyne measurement results. To represent the typical cases, we consider the fidelity of the output state averaged over the measurement outcome.

For a GKP state, we can further improve the performance by utilizing its intrinsic error-correction capability. To do so, in the direct conversion scheme, we apply a quantum limited amplifier with gain $1/\eta_{\rm DC}$ prior to transduction. For the teleportation based scheme, we apply a quantum-limited amplifier with gain $1/\kappa$ prior to transduction when $\kappa<1$, and an attenuator with transmissitivty $1/\kappa$ post transduction when $\kappa>1$. Then, the transduction process is converted to an additive white Gaussian noise channel (see Appendix~\ref{app:additive_noise}). % (see Sec. VII of Ref.~\cite{supp}). %. 
After transduction, one can perform error correction for the GKP state via measuring both quadratures modulus $\sqrt{\pi}$ and performing a displacement to correct the noise. Given an additive noise with variance $\sigma^2$, the leading-order success probability for GKP error correction can be obtained as (see Appendix~\ref{app:GKP}) %(see Sec. VIII of Ref.~\cite{supp})
\begin{align}
P_s = \left[\text{Erf}(\frac{\sqrt{\pi}}{2\sqrt{2(\sigma^2+2\sigma_{\text{GKP}}^2)}})\right]^2,
\end{align}
where $\text{Erf}(x)\equiv (2/\sqrt{\pi})\int_0^x \ dt \exp(-t^2)$ is the Gaussian error function.
%and the square comes from the symmetric success probability along position and momentum directions.

First we consider the ideal case of $\zeta_{\rm m}=1$, $\zeta_{\rm o}=1$.
%, where the noise $n_{\rm in}$ is irrelevant.
We plot the fidelity for the coherent and cat states, and the success probability for GKP states with different cooperativity values in Fig.~\ref{fig:fidelity_performance}(a). The teleportation-based transduction scheme (dashed) provides consistent advantages over the direct transduction scheme (solid) for all three states. As cooperativity ${{C}}$ approaches unity, the fidelity and success probability approach unity and the gap vanishes as expected.

%To understand the performance in a state-of-the-art experimental systems, 
With non-ideal extraction efficiencies ($\zeta_{\rm m}=0.95$, $\zeta_{\rm o}=0.9$) and non-zero noise \QZ{$n_{\rm in}=0.2$}, the GKP success probability with the teleportation scheme (green dashed) is always higher than the direct conversion scheme (green solid), as shown in Fig.~\ref{fig:fidelity_performance}(b). For coherent and cat states, the teleportation scheme offers a better fidelity over the direct conversion when the cooperativity is small. Indeed, at the $C\to 0$ limit, as the direct transmission scheme simply outputs noise independent of the input, while the teleportation scheme always produces output containing some information about the input from the quadrature measurements. When the cooperativity is high, however, direction conversion offers a better fidelity. In particular, at the limit of $C\to 1$ (see Appendix~\ref{app:channel_definition}), 
although the transmissivity of the teleportation-based scheme $\kappa$ can be tuned close to unity, the additive noise mixed in is larger than that of the direct conversion scheme. When $\alpha$ is small, the states are close to vacuum, and a lower transmissivity does not affect the fidelity as much as higher noise, which caused the behavior when $C$ is large. However, when $\alpha$ takes larger values, the fidelity is mainly determined by the transmissivity and the teleportation scheme will offer a better fidelity, as verified in Appendix~\ref{app:fidelity}.

Finally, we vary the extraction efficiencies $\zeta_{\rm m}$ and $\zeta_{\rm o}$ for the practical experimental setting of ${{C}}=0.1$ and \QZ{$n_{\rm in}=0.2$}. As we expect, at relatively low cooperativity, the teleportation based transduction provides a better performance over the direct transduction for all values of extraction efficiencies.
As shown in Fig.~\ref{fig:state}(a)(b), the fidelity of the coherent state is much higher with the teleportation based scheme in (b) compared with the direct conversion in (a). For the cat state, similar advantage can be seen in Fig.~\ref{fig:state}(c)(d), despite the overall fidelity of both schemes to be lower. With the assistance of error correction, the success probability of GKP states transfer is much higher, as shown in Fig.~\ref{fig:state} (e)(f). And the teleportation based transduction provides a much higher success probability. 
%Note that similar advantages from teleportation hold for the different values of squeezing noise $\sigma_{\rm GKP}$, while the overall fidelity is higher with small $\sigma_{\rm GKP}$. 

\section{Discussion} We have proposed a microwave-optical transduction scheme based on continuous-variable teleportation. The scheme overcomes the low-cooperativity obstacle in the direct conversion scheme and provides appreciable advantages in the quantum capacity and state transfer performance. While the analysis is based on cavity electro-optics, the scheme also applies to other transduction systems with intermediate excitations~\cite{zhong2020proposal,lau2020ground}. Additional noise introduced by intermediate excitations needs to be considered in that case. Moreover, the teleportation scheme can also be used for quantum state transduction between different optical frequencies, e.g. diamond color centers for quantum memories and telecom band for long-distance communications. In this case, thermal noise vanishes for both input and output frequencies, corresponding to the ideal case in our analysis (Fig.~\ref{fig:capacity}(a) and Fig.~\ref{fig:fidelity_performance}(a)). Therefore, appreciable advantage can be obtained by using teleportation-based transduction scheme regardless of device cooperativity and extraction efficiency. In terms of applications, beside error correction, our transduction scheme can be applied to sensing protocols like microwave quantum illumination~\cite{barzanjeh2015microwave} and ranging~\cite{zhuang2021quantum_ranging}.
%Although we only addressed microwave-optical transduction, the theory framework of our protocol can be generalized to transduction systems between other frequencies, as long as the interaction can be configured to beamsplitter and two-mode-squeezing forms.

\begin{acknowledgments}
This research project is supported by the Defense Advanced Research Projects Agency (DARPA) under Young Faculty Award (YFA) Grant No. N660012014029, Office of Naval Research Grant No. N00014-19-1-2189 and Office of Naval Research Grant No. N00014-19-1-2190.
\end{acknowledgments}

%\bibliography{Ref.bib}


\appendix

\section{Preliminary}
\label{app:preliminary}
\subsection{Wigner function}
Below we introduce the basic notations and definitions, which are similar to the formalism in Ref.~\cite{Nohsurface-GKP}. Readers can also refer to Ref.~\cite{Weedbrook_2012} for a review of Gaussian quantum information.
Let $\calH$ denote an infinite-dimensional Hilbert space. We consider $n$ bosonic modes associated with tensor product of $n$ Hilbert space $\calH^{\otimes n}$, which have $n$ pairs of independent annihilation and creation operators $(\hat{a}_k,\hat{a}^\dagger_k)$ satisfying $[\hat{a}_i,\hat{a}_j^\dagger]=\delta_{i j}$. We define the quadrature operators as  $\hat{q}_k \equiv (\hat{a}_k+\hat{a}^\dagger_k)/\sqrt{2}$, $\hat{p}_k \equiv i (\hat{a}^\dagger_k-\hat{a}_k)/\sqrt{2}$ for unit $\hbar=1$. The vector quadrature operator is defined as $\hat{\bx} \equiv (\hat{q}_1,\hat{p}_1,...,\hat{q}_n,\hat{p}_n)^{\rm T}$ which satisfies the commutation relation $[\hat{x}_i,\hat{x}_j]=i\bOmega_{i j}$, with matrix
\begin{equation}
    \bOmega \equiv \bigoplus_{k=1}^n \omega, \; \omega = \begin{pmatrix}
    0 & 1\\
    -1 & 0\\
    \end{pmatrix}.
\end{equation}
An $n$-mode displacement operator $\hat{D}(\balpha)$ is defined as $\hat{D}(\balpha)\equiv \exp(\balpha^{\rm T}\hat{\bolda}^\dagger-(\balpha^*)^{\rm T}\hat{\bolda})$, where $\balpha=(\alpha_1,\dots,\alpha_n)^{\rm T}$ is a vector of $n$ complex numbers and $\hat{\bolda}=(\hat{a}_1,\dots,\hat{a}_n)^{\rm T}$, $\hat{\bolda}^\dagger=(\hat{a}_1^\dagger,\dots,\hat{a}_n^\dagger)^{\rm T}$. Alternatively, one can express the displacement operator in the Weyl form $\hat{D}(\balpha) =\exp(i \hat{\bx} ^{\rm T}\bOmega \bxi)\equiv \hat{D}(\bxi)$, where the $2n$ dimensional vector $\bxi\equiv\sqrt{2}(\rm{Re}(\alpha_1),\rm{Im}(\alpha_1),\dots,\rm{Re}(\alpha_n),\rm{Im}(\alpha_{n}))$. The Weyl operator satisfies:
\begin{align} 
    &\text{Tr}[\exp(-i \hat{\bx} ^{\rm T}\bOmega \bxi_1)\exp(i \hat{\bx} ^{\rm T}\bOmega \bxi_2)]=(2\pi)^n \delta(\bxi_1-\bxi_2),\\
    & \int d^{2n} \bxi \; \exp(i \hat{\bx} ^{\rm T}\bOmega \bxi) \hat{A}  \exp(- i \hat{\bx} ^{\rm T}\bOmega \bxi) = (2 \pi)^n \text{Tr}(\hat{A})\hat{I}.
\end{align}
With the above relations, we have the following transform pairs:
\begin{align}
    & \chi(\bxi ; \hat{\rho}) \equiv \text{Tr}[\hat{\rho} \exp(i \hat{\bx} ^{\rm T}\Omega \bxi)],\\
    & \hat{\rho} = \frac{1}{(2\pi)^n} \int d^{2n} \bxi \; \chi(\bxi;\hat{\rho}) \exp(-i \hat{\bx}^{\rm T}\bOmega \bxi),\\
    & W(\bx;\hat{\rho})=\frac{1}{(2\pi)^n} \int d^{2n}\bxi \; \chi(\bxi;\hat{\rho}) \exp(-i \bx^{\rm T}\bOmega \bxi),\\
    & \chi(\boldsymbol{\xi};\hat{\rho})=\frac{1}{(2\pi)^n} \int d^{2n}\bx \; W(\bx;\hat{\rho}) \exp(i \bx^{\rm T}\bOmega \bxi),
\end{align}
where $\chi(\bxi ; \hat{\rho})$ and $W(\bx;\hat{\rho})$ are the characteristic function and Wigner function of state $\hat{\rho}$. 

The following are well-known properties of Wigner function that we will utilize in our calculations:
\begin{align}
    & \int d^{2n} \bx \; W(\bx;\hat{\rho})=1, \label{eq:Wigner-normalization}\\
    & W(\bx;\hat{\rho}_1\otimes\hat{\rho}_2)=W(\bx_1;\hat{\rho}_1)W(\bx_2;\hat{\rho}_2),
    \label{eq:Wigner-direct-product}\\
    & \text{Tr}(\hat{\rho}\hat{\sigma})= (2\pi)^n \int d^{2n} \bx \; W(\bx;\hat{\rho})W(\bx;\hat{\sigma}),
    \label{eq:Wigner-inner-product}\\
    &W\left(\bx_A;\text{Tr}_B (\hat{\rho}_{AB})\right) = \int d^2 \bx_B \; W(\bx_A, \bx_B; \hat{\rho}_{AB}).
    \label{eq:Wigner-Trace}
\end{align}

\subsection{Gaussian states and unitaries}

A quantum state is Gaussian if its Wigner function $W(\bx)$ has the Gaussian form
\begin{equation}
    W(\bx) = \frac{1}{(2\pi)^n \vert\boldsymbol{V}\vert^{\frac{1}{2}}} \exp[-\frac{1}{2}(\bx - \Bar{\bx})^{\rm T}\boldsymbol{V}^{-1}(\bx-\Bar{\bx})],
    \label{eq:Wigner-Gaussian-state}
\end{equation}
where $\Bar{\bx}$ and $\boldsymbol{V}$ are the mean and covariance matrix of state $\hat{\rho}$:
\begin{align}
    & \Bar{\bx} \equiv \text{Tr}[\hat{\rho} \hat{\bx}],\\
    & V_{ij} \equiv
    \frac{1}{2}\text{Tr}[\hat{\rho} \{\hat{x}_i-\bar{x}_i,\hat{x}_j-\bar{x}_j\}].
\end{align}
Here we give two examples of Gaussian state relevant to our calculations. A thermal state has $\Bar{\bx}=\boldsymbol{0}$, $\boldsymbol{V} = {(1/2+N)} \bold{I}_2$, where $N$ is the mean photon number of the thermal state and $\bold{I}_k$ is a $k$ by $k$ identity matrix. A vacuum state $\ket{0}$ is a thermal state with $N=0$. A coherent state is defined by displacing the vacuum state as $\ket{\alpha} \equiv \hat{D}(\alpha)\ket{0}$. It is also a Gaussian state with $\Bar{\bx}=\sqrt{2}\boldsymbol{\alpha}$, $\boldsymbol{V} = \bold{I}_2/2$.

A Gaussian unitary is generated by a Hamiltonian in a second-order polynomial of $\hat{\bolda}$ and $\hat{\bolda}^\dagger$. It is fully characterized by a symplectic matrix $\boldsymbol{S}$ and a vector $\boldsymbol{d}$. Thus we denote it as $\hat{U}_{\boldsymbol{S},\boldsymbol{d}}$. In the Heisenberg picture, it transforms the vector quadrature operator linearly via
\begin{equation}
    \hat{\bx} \rightarrow \boldsymbol{S} \hat{\bx} + \boldsymbol{d}.
\end{equation}
A Gaussian state is mapped to another Gaussian state under the transform $\hat{U}_{\boldsymbol{S},\boldsymbol{d}}$, with the mean and covariance matrix transformed as
\begin{equation}
    \bar{\bx}\rightarrow\boldsymbol{S}\bar{\bx}+\boldsymbol{d},\; \boldsymbol{V}\rightarrow\boldsymbol{S}\boldsymbol{V}\boldsymbol{S}^{\rm T}.
\end{equation}
One can also show that, for any quantum state $\hat{\rho}$, its Wigner function transforms as
\begin{equation}
     W(\bx^\prime;\hat{U}_{\bold{S},\bold{d}}\hat{\rho}\hat{U}^\dagger_{\bold{S},\bold{d}}) = W\left[\bold{S}^{-1}(\bx^\prime-\bold{d});\rho\right]
     \label{eq:Wigner-Gaussian-transform}
\end{equation}
under the Gaussian unitary.

\section{Analysis of the interaction Hamiltonian}
\label{app:generation}

Inside the cavity, the $\chi^{(2)}$ nonlinear material immersing in a strong classic pump field will proceed typical interaction between the optical and microwave fields. In general, a spontaneous parametric down-conversion (SPDC) process will be triggered when the pump frequency is equal to the sum frequency of the optical and microwave fields. While if the pump frequency matches the frequency differences between the two fields, the interaction will act as a frequency-domian beamsplitter.

As depicted in Fig.~\ref{fig:schematic}, the entanglement between the optical and microwave fields can be generated from an SPDC process by pumping a triple-resonance device \cite{fan2018superconducting}. This electro-optical system with coupling strength $g_E$ is modelled by the total Hamiltonian
\begin{align}
    H=\hbar \omega_o \hat{a}^{\dagger}\hat{a} + \hbar \omega_m \hat{m}^{\dagger}\hat{m} + i\hbar (g_E\hat{a}^{\dagger}\hat{m}^{\dagger}-g_E^*\hat{a}\hat{m}),
\end{align}
where $\hat{a}$ ($\hat{m}$) is the annihilation operator for optical (microwave) field with resonance at frequency $\omega_o$ ($\omega_m$). The intra-cavity pump power and the phase-matching condition have been absorbed into the coupling strength $g_E$.

The output fields can be derived by solving a group of Heisenberg-Langevin equations in the Fourier domain with the input–output relations \cite{zhong2020proposal,cui2021high}
\begin{align}
    &0=\bold{G}\abold + \bold{K}\abold_{\rm {in}},\\
    &\abold_{\rm {out}}=-\bold{K}^{\rm T}\abold+\abold_{\rm {in}}.
\end{align}
Here we use the matrix form to represent the dynamics in the Fourier domain with optical frequency detuning $\Delta_p=\omega-\omega_o$ and microwave frequency detuning $\Delta_e=\omega-\omega_m$. The notations are defined as the following
\begin{align}
    &\abold=(\hat{a}, \hat{a}^\dagger, \hat{m}, \hat{m}^\dagger  )^{\rm T},\\
    &\abold_{\rm {in}}=(\hat{a}_{\rm {in}}, \hat{a}^\dagger_{\rm {in}},\hat{a}^{(i)}, \hat{a}^{\dagger(i)}, \hat{m}_{\rm {in}}, \hat{m}^\dagger_{\rm {in}},\hat{m}^{(i)}, \hat{m}^{\dagger(i)}  )^{\rm T},\\
    &\bold{G}=\begin{pmatrix}
    -\frac{\gamma_p}{2}+i\Delta_p & 0 & 0 & -ig_E\\
    0 & -\frac{\gamma_p}{2}-i\Delta_p & ig_E & 0\\
    0 & -ig_E & -\frac{\gamma_e}{2}+i\Delta_e & 0\\
    ig_E & 0 & 0 & -\frac{\gamma_e}{2}-i\Delta_e
    \end{pmatrix},\\
    &\bold{K}=\begin{pmatrix}
    \sqrt{\gamma_{pc}}& 0 &\sqrt{\gamma_{pi}} & 0 & 0 & 0 & 0 & 0 \\
    0 & \sqrt{\gamma_{pc}}& 0 &\sqrt{\gamma_{pi}} & 0 & 0 & 0 & 0\\
    0 & 0 & 0 & 0 & \sqrt{\gamma_{ec}}& 0 &\sqrt{\gamma_{ei}} & 0\\
    0 & 0 & 0 & 0 & 0 & \sqrt{\gamma_{ec}}& 0 &\sqrt{\gamma_{ei}}
    \end{pmatrix}.
\end{align}
The output fields then relate to the input fields by 
\begin{align}
    \abold_{\rm {out}}=\bold{S}_a \abold_{\rm {in}}=(\bold{K}^{\rm T}\bold{G}^{-1}\bold{K}+\bold{I}_8)\abold_{\rm {in}}.
\end{align}
By defining quadrature observables and the transform matrix as
\begin{align}
    \begin{pmatrix}
    \hat{q}^{a}\\
    \hat{p}^{a}
    \end{pmatrix}&=\frac{1}{\sqrt{2}}\begin{pmatrix}
    1 & 1\\
    -i & i
    \end{pmatrix}\begin{pmatrix}
    \abold\\
    \abold^\dagger
    \end{pmatrix},\\
    \bold{Q}&=\bold{I}_4\otimes\frac{1}{\sqrt{2}}\begin{pmatrix}
    1 & 1\\
    -i & i
    \end{pmatrix},
\end{align}
we get the input-output quadrature relation
\begin{align}
    &\hat{\bold{x}}_{\rm {out}}=\bold{S}_x\hat{\bold{x}}_{\rm {in}}=\bold{Q}\bold{S}_a\bold{Q}^{-1}\hat{\bold{x}}_{\rm {in}},\\
    &\hat{\bold{x}}_{\rm {in}} = (\hat{q}^{p}_{\rm {in}}, \hat{p}^{p}_{\rm {in}},\hat{q}^{p,(i)}, \hat{p}^{p,(i)}, \hat{q}^{e}_{\rm {in}}, \hat{p}^{e}_{\rm {in}},\hat{q}^{e,(i)}, \hat{p}^{e,(i)}  )^{\rm T}.
\end{align}
Then, the input-output relation of the covariance matrix is derived as
\begin{align}
    \bold{V}_{\rm {out}}=\bold{S}_x\bold{V}_{\rm {in}}\bold{S}_x^{\rm T},
\end{align}
in which the input covariance matrix $\bold{V}_{\rm {in}}$ contains vacuum noise from the optical modes and the input microwave mode while the dissipation microwave mode is contaminated by the thermal noise of population $n_{\rm {in}}$,
\begin{align}
    \bold{V}_{\rm {in}}={\rm Diag}(\bold{I}_{6},(n_{\rm {in}}+1/2)\bold{I}_{2}).
\end{align}
Taking the input covariance matrix into consideration, the covariance matrix of two output fields finally shows in the form as
\JW{
\begin{align}
    V_{\rm{o,m}}=\frac{1}{2}\begin{pmatrix}
    w& 0 & 0 & -v\\
    0 & w &-v & 0\\
    0 & -v & u & 0\\
     -v & 0 & 0 & u
    \end{pmatrix},
    \label{eq:CM_optical-electric}
\end{align}
}

\begin{subequations}
where we reprint the parameters in Eq.~\eqref{uvw} of the main paper as follows,
\JW{
\begin{align}
&u=1+\frac{8\zeta_{\rm m} [{{C}}+ n_{\rm in}(1-\zeta_{\rm m})]}{(1-{{C}})^2},
\\
&v=\frac{4\sqrt{\zeta_{\rm o}\zeta_{\rm m} {{C}}}[1+{{C}}+2 n_{\rm in}(1-\zeta_{\rm m})]}{(1-{{C}})^2},
\\
&w=1+\frac{8{{C}}\zeta_{\rm o}\left[1+n_{\rm in}\left(1-\zeta_{\rm m}\right)\right]}{(1-{{C}})^2}.
\end{align}
}
\label{uvw_app}
\end{subequations}

\section{Continuous-variable teleportation}
\label{app:teleportation}

\begin{figure}
    \centering
    \includegraphics[width=0.45\textwidth]{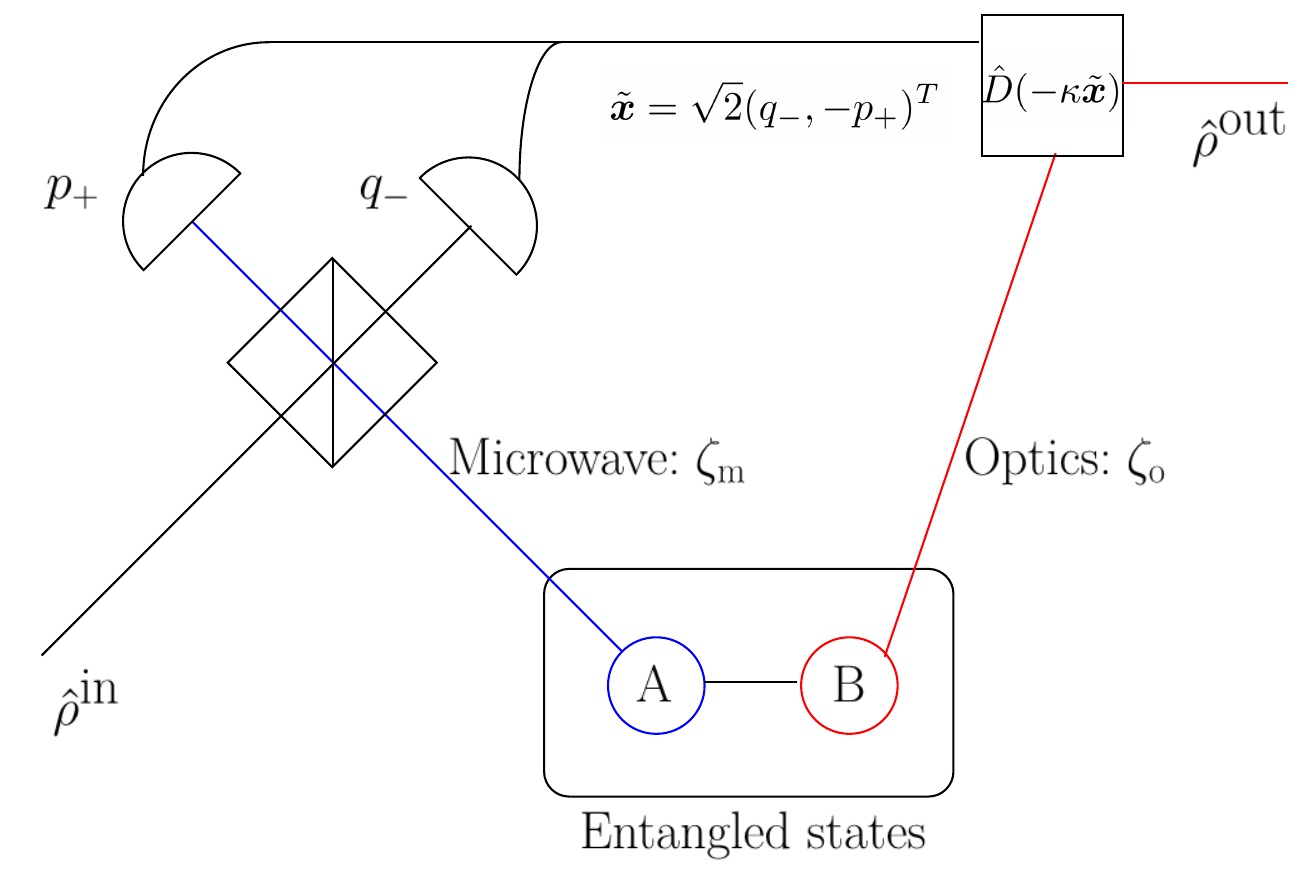}
    \caption{Schematic of the continuous-variable teleportation. }
    \label{fig:schematic_teleportation}
\end{figure}

%$\hat{a}_1\rightarrow e^{i\pi/2}\hat{a}_1$

Here we consider teleportation utilizing the entangled state  $\hat{\rho}_{AB}$ between the microwave domain (A) and optical domain (B) as shown in Fig.~\ref{fig:schematic_teleportation}. To make the analyses more convenient,\JW{ we apply a $\pi/2$-phase rotation to the microwave mode to convert the covariance matrix of (\ref{eq:CM_optical-electric}) to a standard form
\begin{align}
&V_{\rm{m,o}} = \frac{1}{2}
\begin{pmatrix}
u \bI_2 & v \bZ_2 \\
v \bZ_2 & w\bI_2 
\end{pmatrix},
\label{eq:CM_app}
\end{align}
}
where $\bZ_2 = \mbox{Diag} (1, -1)$ is the Pauli-Z matrix. \QZ{Note that we have re-ordered the microwave mode as the first mode and the optical mode as the second mode above, compared with the covariance matrix of Eq.~\eqref{eq:CM_optical-electric}}

As shown in Fig.~\ref{fig:schematic_teleportation}, in a continuous-variable teleportation scheme, to transmit an input state $\hat{\rho}^{\rm in}$ from the microwave domain to the optical domain, one performs a balanced beamsplitter on the microwave subsystem $A$ of the entangled state and the microwave input state $\hat{\rho}^{\rm in}$. Then the position quadrature $\hat{q}_-$ and the momentum quadrature $\hat{p}_+$ of the beamsplitter outputs are measured by homodyne. The rescaled measurement result $\tilde{\bm x}=\sqrt{2}(q_-,-p_+)$ is then utilized to perform a displacement of amount $-\kappa \tilde{\bm x}$ on the optical subsytem $B$ of the entangled state to produce the optical output $\hat{\rho}^{\rm out}$. Here we have added a factor of $\sqrt{2}$ in $\tilde{\bm x}$ because the balanced beamsplitter introduces a $1/\sqrt{2}$ factor in quadratures.

In the following, we present the detailed calculation of the Wigner function of the output. We label the pre-shared entangled state's Wigner function as $W^{AB}(\bx_A,\bx_B)$ and input state's Wigner functions as $W^{\rm {in}}(\bx_{\rm {in}})$. The Wigner function of system $AB$ and input is a direct product $W^{\rm {in}}(\bx_{\rm {in}})W^{AB}(\bx_A,\bx_B)$. Then a $50/50$ beam splitter (BS) is applied to transform the quadeatures
\begin{equation}
    \begin{pmatrix}
    \bx_{+}\\
    \bx_{-}
    \end{pmatrix}
    =\frac{1}{\sqrt{2}}
    \begin{pmatrix}
    \bI_2 & \bI_2\\
    -\bI_2 & \bI_2
    \end{pmatrix}
    \begin{pmatrix}
    \bx_{\rm {in}}\\
    \bx_{A}
    \end{pmatrix}.
\end{equation}
The transform of Wigner function is obtained from Eq.~\eqref{eq:Wigner-Gaussian-transform}: $W^{\rm {in}}(\frac{\bx_{+}-\bx_{-}}{\sqrt{2}})W^{AB}(\frac{\bx_{+}+\bx_{-}}{\sqrt{2}},\bx_B)$.
Then one gets measurement results from homodyne detection $\tilde{\bx}=\sqrt{2}(q_{-},-p_{+})^{\rm T}$. At Bob's side, the Wigner function of system B is obtained by tracing out $q_{+}$ and $p_{-}$. Thus,
\begin{align*}
    W(\bx_B \vert \tilde{\bx}) =& c \int \diff{q_{+}}\diff{p_{-}} W^{\rm {in}}(\frac{q_{+}-q_{-}}{\sqrt{2}},\frac{p_{+}-p_{-}}{\sqrt{2}}) \\
    & \quad \times W^{AB}(\frac{q_{+}+q_{-}}{\sqrt{2}},\frac{p_{+}+p_{-}}{\sqrt{2}},\bx_B),
\end{align*}
where $c$ is a constant that normalizes $W(\bx_B\vert\tilde{\bx})$.
With the substitution of $\bx = (\frac{q_{+}-q_{-}}{\sqrt{2}},\frac{p_{+}-p_{-}}{\sqrt{2}})^{\rm T}$, $W(\bx_B \vert \tilde{\bx})$ can be written as
\begin{equation}
    W(\bx_B\vert\tilde{\bx})= c \int d^2 \bx \; W^{AB}\left[\bZ_2(\bx + \tilde{\bx}),\bx_B\right] W^{\rm {in}}(\bx).
\end{equation}
Suppose we perform a displacement $\hat{D}(-\kappa \tilde{\bx})$ at Bob's side to recover the input state, then the Wigner function of the output state is obtained from property \eqref{eq:Wigner-Gaussian-transform} as
\begin{align}
        &W^{\rm {out}}(\bx_B\vert\tilde{\bx})= \nonumber\\
        &\quad c \int d^2 \bx \; W^{AB}\left[\bZ_2(\bx + \tilde{\bx}),\bx_B+\kappa \tilde{\bx}\right] W^{\rm {in}}(\bx).
        \label{eq:Wigner-Bobside}
\end{align}
To understand the recovery of the input state, we first consider the ideal infinite entangled limit, where the Wigner function $W^{AB}(\bx_A,\bx_B)=\delta(\bx_A-\bZ_2 \bx_B)$. In this ideal case, we may take $\kappa = 1$ and the output Wigner function 
\begin{align*}
    W^{\rm {out}}(\bx_B)&= \int d^2 \bx \delta\left[\bZ_2(\bx-\bx_B)\right] W^{\rm {in}}(\bx)= W^{\rm {in}}(\bx_B).
\end{align*}
is independent of $\tilde{\bx}$. However this may not be true for the general Gaussian entangled state. 
To prepare our calculation, we note that the inverse of the \QZ{matrix} in Eq.~\eqref{eq:CM_app} can be written as:
\begin{equation}
    V_{\QZ{\rm m,o}}^{-1}= \frac{2}{uw-v^2}
    \begin{pmatrix}
    w \bI_2 & -v \bZ_2 \\
    -v \bZ_2 & u \bI_2 
    \end{pmatrix}.
\end{equation}
We consider $W^{AB}$ given by Eq.~\eqref{eq:Wigner-Gaussian-state} with a zero mean and the above $V^{-1}$. Then Eq.~\eqref{eq:Wigner-Bobside} gives
\begin{align}
&W^{\rm {out}}(\bx_B\vert\tilde{\bx})= c \; \exp[-\frac{1}{w}(\bx_B+\kappa\tilde{\bx})^2] \int d^2 \bx
\\
&
 \; \exp[-\frac{w}{u w -v^2}\left(\bx-\frac{v}{w}\bx_B+(1-\frac{v}{w}\kappa)\tilde{\bx}\right)^2] W^{\rm {in}}(\bx).
\end{align}

We can obtain the normalize constant $c$ from Eq.~\eqref{eq:Wigner-normalization} as
\begin{equation}
    c = \frac{u}{\pi(u w - v^2)}\frac{1}{\int d^2 \bx \; \exp\left[-\frac{1}{u}(\bx+\tilde{\bx})^2\right] W^{\rm {in}}(\bx)}.
\end{equation}
The probability density function (PDF) of getting the results $\tilde{\bx}$ is given by:
\begin{align}
    \nonumber
    P(\tilde{\bx}) & = \int \diff{q_{+}}\diff{p_{-}} \; W^A(\frac{\bx_{+}+\bx_{-}}{\sqrt{2}}) W^{\rm {in}}(\frac{\bx_{+}-\bx_{-}}{\sqrt{2}})\\
    \nonumber
    & =  \int d^2\bx \; W^A[\bZ_2(\bx+\tilde{\bx})] W^{\rm {in}}(\bx)\\
    & = \frac{1}{\pi u } \int d^2\bx \; \exp\left[-\frac{1}{u}(\bx+\tilde{\bx})^2\right] W^{\rm {in}}(\bx).
\end{align}
The average state is then
\begin{widetext}
\begin{align}
    \nonumber
    W^{\rm {out}}(\bx_B)& = \int d^2\tilde{\bx} \;  W^{\rm {out}}(\bx_B\vert\tilde{\bx}) P(\tilde{\bx})\\
    & = \frac{1}{\pi(u \kappa^2 - 2v\kappa+w)} \int d^2\bx \; W^{\rm {in}}(\bx) \exp[-\frac{\kappa^2}{u \kappa^2 - 2v\kappa+w}(\bx-\frac{1}{\kappa}\bx_B)^2].
   \label{eq:teleported-average-state}
\end{align}
\end{widetext}

\section{Thermal-amplifier and thermal-attenuator channels}
\label{app:channel_definition}

\subsection{Wigner function transforms}
\begin{figure}[t]
\centering
\includegraphics[width=0.35\textwidth]{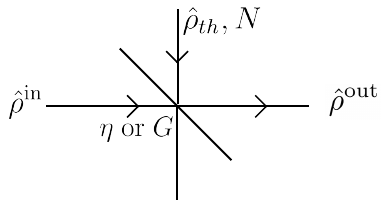}
\caption{Schematic of an thermal-attenuator/amplifier channel. The thermal state $\hat{\rho}_{th}$ has mean photon number $N$. $\eta$ is the attenuator transmissivity and $G$ is the amplifier gain.
\label{fig:Model}
}
\end{figure}

In this section, we first derive the Wigner function input-output relation for thermal-amplifier and thermal-attenuator channels, as shown in Fig.~\ref{fig:Model}. Then, we show that the quantum teleportation process reduces to a thermal-amplifier or a thermal-attenuator, via comparing the results with Eq.~\eqref{eq:teleported-average-state}. 

We assume that the Wigner function of input state is $W^{\rm {in}}(\bx)$, the other input is a thermal state with mean photon number $N$ and the Wigner function of output state is $W^{\rm {out}}(\bx_B)$. By utilizing the Gaussian unitary transform property in Eq.~\eqref{eq:Wigner-Gaussian-transform}, we obtain for the thermal-attenuator channel $\calL_{\eta,N}$ with transmissivity $\eta$ and noise $N$
\begin{equation}
    W^{\rm {out}}(\bx_B) \propto\int d^2\bx \; W^{\rm {in}}(\bx) \exp[-\frac{(\bx-\frac{1}{\sqrt{\eta}}\bx_B)^2}{(1+2N)(1-\eta)/\eta}].
    \label{wigner;thermal_attenuator}
\end{equation}
% where $c = \frac{1}{\pi(1+2N)(1-\eta)} $.
While for the thermal-amplifier $\calA_{G,N}$ with gain $G$ and noise $N$
\begin{equation}
    W^{\rm {out}}(\bx_B) \propto \int d^2\bx \; W^{\rm {in}}(\bx) \exp[-\frac{(\bx-\frac{1}{\sqrt{G}}\bx_B)^2}{(1+2N)(G-1)/G}].
\end{equation}
Now the channel model of the teleportation can be obtained from comparing the above with Eq.~\eqref{eq:teleported-average-state}. If $\kappa<1$, the continuous-variable teleportation channel $\calT$ reduces to a thermal-attenuator channel $\calL_{\eta,N}$ with:
\begin{align}
    & \eta= \kappa^2,\\
    & 1+2N =\frac{u \kappa^2-2v\kappa + w}{1-\kappa^2}.
\end{align}
Otherwise, the continuous-variable teleportation channel $\calT$ may be modeled as a thermal-amplifier $\calA_{G,N}$ with:
\begin{align}
    & G = \kappa^2,\\
    & 1+2N = \frac{u \kappa^2-2v\kappa + w}{\kappa^2 - 1}.
\end{align}
When $\kappa=1$, it is an additive noise channel with the noise variance $u\kappa^2-2v\kappa+w$.

The above can be summarized as
\begin{align}
\calT_{\kappa^2,N_{\rm TP}}=
\left\{ \begin{array}{ll}
         \calL_{\kappa^2,N_{\rm TP}} & \mbox{if $\kappa\le 1$};\\
        \calA_{\kappa^2,N_{\rm TP}} & \mbox{if $\kappa>1$},\end{array} \right.
\end{align}
with $N_{\rm TP}$ defined in Eq.~\eqref{N_TP}; the full expression in terms of $\zeta_{\rm o}$, $\zeta_{\rm m}$ and $C$ is given by:
% \begin{widetext}
% \begin{align}
%     \nonumber
%     & \abs{1-\kappa^2}(1+2N_{\rm TP}) =
%     \\
%     & \nonumber 
%     \quad
%     1+\kappa^2+\frac{8}{(1-C)^2} \left\{\left[1+n_{\rm in}\left(1-\zeta_{\rm m}\right)\right]\left[\zeta_{\rm m}-2\kappa\sqrt{C\zeta_{\rm o}\zeta_{\rm m}}+\kappa^2\zeta_{\rm o}\right] \right.\\
%     & \left.
%     \quad +
%      \left(1-C\right)
%      \left[-\zeta_{\rm m}+\kappa\sqrt{C\zeta_{\rm o}\zeta_{\rm m}}-\kappa^2\left(1+n_{\rm in}\left(1-\zeta_{\rm m}\right)\right)\zeta_{\rm o}\right]\right\}.
%      \label{eq:Ntp-full-expression}
% \end{align}
% \end{widetext}
\JW{
\begin{widetext}
\begin{align}
    \nonumber
    & \abs{1-\kappa^2}(1+2N_{\rm TP}) =
    \\
    & \nonumber 
    \quad
    1+\kappa^2+\frac{8}{(1-C)^2} \left\{\left[1+n_{\rm in}\left(1-\zeta_{\rm m}\right)\right]\left[\zeta_{\rm o}-2\kappa\sqrt{C\zeta_{\rm o}\zeta_{\rm m}}+\kappa^2\zeta_{\rm m}\right] \right.\\
    & \left.
    \quad +
     \left(1-C\right)
     \left[-\zeta_{\rm m}\kappa^2+\kappa\sqrt{C\zeta_{\rm o}\zeta_{\rm m}}-\left(1+n_{\rm in}\left(1-\zeta_{\rm m}\right)\right)\zeta_{\rm o}\right]\right\}.
     \label{eq:Ntp-full-expression}
\end{align}
\end{widetext}
}
In the limit of $C \rightarrow 1$, the first term in the bracket diverges unless \JW{ $\zeta_{\rm o}-2\kappa\sqrt{C\zeta_{\rm o}\zeta_{\rm m}}+\kappa^2\zeta_{\rm m}=0$.} So we need \JW{$\kappa=\sqrt{\zeta_{\rm o}/\zeta_{\rm m}}$.} Substituting $\kappa$ into Eq.~\eqref{eq:Ntp-full-expression}, it simplifies to 
% \begin{align}
%     & \nonumber
%     \abs{1-\kappa^2}(1+2N_{\rm TP}) = \\
%     & 1+\frac{\zeta_{\rm m}}{\zeta_{\rm o}}-\frac{8\zeta_{\rm m}\left[\sqrt{C}-(1-\zeta_{\rm m})n_{\rm {in}}\right]}{(1+\sqrt{C})^2}.
% \end{align}
\JW{
\begin{align}
    & \nonumber
    \abs{1-\kappa^2}(1+2N_{\rm TP}) = \\
    & 1+\frac{\zeta_{\rm o}}{\zeta_{\rm m}}+\frac{8\zeta_{\rm o}\left[1+(1-\zeta_{\rm m})n_{\rm {in}}\right]}{(1+\sqrt{C})^2}.
\end{align}
}
When $C=1$, we finally get
% \begin{equation}
%     \abs{1-\kappa^2}(1+2N_{\rm TP}) =1+\zeta_{\rm m}\left(\frac{1}{\zeta_{\rm o}}+2n_{\rm in}(1-\zeta_{\rm m})-2\right).
% \end{equation}
\JW{
\begin{equation}
    \abs{1-\kappa^2}(1+2N_{\rm TP}) =1+\zeta_{\rm o}\left(\frac{1}{\zeta_{\rm m}}+2n_{\rm in}(1-\zeta_{\rm m})+2\right).
\end{equation}
}
For direct conversion, at the high cooperativity limit ($C\to 1$), from Eqs.~\eqref{eta_DC} and \eqref{N_DC} of the main paper we have
$
\eta_{\rm DC}=\zeta_{\rm m} \zeta_{\rm o}
$
and
$
(1-\eta_{\rm DC})N_{\rm DC}=(1-\zeta_{\rm m}) \zeta_{\rm o} n_{\rm in}.
$
With the parameters in Fig.~\ref{fig:fidelity_performance}(b), we have $\kappa\simeq 0.973$, $\abs{1-\kappa^2}(1+N_{\rm TP})\simeq \JW{3.93}$ for teleportation and $\eta_{\rm DC}=0.86,(1-\eta_{\rm DC})N_{\rm DC}=0.09$ for direct conversion.

When $C=0$, we finally get
\JW{
\begin{equation}
    \abs{1-\kappa^2}(1+2N_{\rm TP}) =1+\kappa^2\left[1+8n_{\rm in}(1-\zeta_{\rm m})\zeta_{\rm m}\right],
\end{equation}
}
while $\eta_{\rm DC}=0$ and $N_{\rm DC}=0$. In this limit, in the direct transmission case the output is entirely independent of the input as the transmissivity is zero. In the teleportation case, the protocol reduces to performing a heterodyne measurement on the input and displace accordingly. Despite the large noise in the output, the teleportation scheme has an output classically correlated to the input, and therefore has better fidelity.

\subsection{Concatenation of a thermal-amplifier and a thermal-attenuator}
\label{app:concatenation}

In this section, we address how to transform a thermal-amplifiers/attenuator to an additive white Gaussian noise channel described by 
\begin{equation}
    \calN_{\sigma^2}(\hat{\rho}) \equiv \frac{1}{\pi \sigma^2}\int \diff^2\alpha\;e^{-\frac{\abs{\alpha}^2}{\sigma^2}} \hat{D}(\alpha)\hat{\rho}\hat{D}^\dagger(\alpha).
\end{equation}
where $\sigma^2$ is the noise variance, via the approach in Ref.~\cite{noh2019quantumcapacity}.
For a thermal-attenuator, we apply an amplifier before the channel
\begin{equation}
    \calL_{\eta,\bar{n}_{th}} \cdot \calA_{1/\eta, 0} = \calN_{\sigma^2(\eta,\bar{n}_{th})},
\end{equation}
where
$
    \sigma^2(\eta,\bar{n}_{th}) \equiv \left(1-\eta \right)\left(\bar{n}_{th}+1\right)
$.
For a thermal-amplifier, we append a pure loss channel afterwards
\begin{equation}
    \calL_{1/G,0} \cdot \calA_{G, \bar{n}_{th}} = \calN_{\sigma^2(G,\bar{n}_{th})},
\end{equation}
where
$
    \sigma^2(G,\bar{n}_{th}) \equiv \left(1-1/G \right)\left(\bar{n}_{th}+1\right)
$.

\section{Bounds on quantum capacity} 
\label{app:q_bound}

For direct conversion, we hope to evaluate the quantum capacity of the thermal-attenuator channel with parameters in Eqs.~\eqref{eta_DC} and~\eqref{N_DC} of the main paper. For teleportation, we hope to evaluate the quantum capacity maximized over $\kappa$. Depending on the choice of $\kappa$, the channel is either a thermal-attenuator or thermal-amplifier.

The quantum capacity lower bound of direct conversion 
\be 
Q_{\rm DC}^{(\rm LB)}=Q_{\rm LB}(\eta_{\rm DC},N_{\rm DC})
\ee
and the quantum capacity lower bound of teleportation 
\be 
Q_{\rm TP}^{(\rm LB)}=\max_{\kappa} Q_{\rm LB}(\kappa^2,N_{\rm TP})
\ee
have the same functional form~\cite{holevo2001evaluating}
\begin{align} 
Q_{\rm LB}(k,N)\equiv  \max\left[\log\left(\frac{k}{|1-k|}\right)-g\left(N\right),0\right],
\label{eq:Qlowerbound}
\end{align}  
where the function
$ 
g(x)=(x+1)\log_2(x+1)-x\log_2 x
$
is the von Neumann entropy of a thermal state with mean occupation number $x$. Note that when $k\to 1$ and $(1-k)N\to N_{\rm add}$, we have
\be 
Q_{\rm LB}(k,N)=-\log_2(N_{\rm add})-1/\ln(2).
\ee 

We will utilize upper bounds derived from two-way assisted quantum capacity~\cite{pirandola2017fundamental} and the degradable extensions~\cite{fanizza2021estimating}. Combining the thermal-attenuator and thermal-amplifier results, we have
\be 
Q_{\rm DC}^{\rm (UB)}=\min[Q_{\rm PLOB}(\eta_{\rm DC},N_{\rm DC}), Q_{\rm DE}(\eta_{\rm DC},N_{\rm DC}) ].
\ee 
Here the functions
\begin{widetext}
\begin{align}
Q_{\rm PLOB}(\eta, N)\equiv 
\left\{ \begin{array}{ll}
         \max\left[-\log_2\left[\left(1-\eta\right)\eta^N\right]-g\left(N\right),0\right] 
         & \mbox{if $\eta< 1$};
         \\
        \max\left[\log_2\left(\frac{\eta^{N+1}}{\eta-1}\right)-g(N),0\right] 
        & \mbox{if $\eta>1$},
        \\ 
        \log_2\left(1/N_{\rm add}\right)-1/\ln(2)+N_{\rm add}/\ln(2).
        & \mbox{if $\eta=1, (1-\eta)N\to N_{\rm add}$}
        \end{array} \right.
\end{align}
\begin{align}
Q_{\rm DE}(\eta, N)\equiv 
\left\{ \begin{array}{ll}
         \max\left[\log_2\left(\frac{\eta}{1-\eta}\right)+h\left[\left(1-\eta\right)\left(2N+1\right)+\eta\right]\JW{-} h\left[\eta\left(2 N +1\right)+1-\eta\right],0\right] 
         & \mbox{if $\eta< 1$};\\
         \JW{\mbox{I changed the sign. See equation (32) of \cite{fanizza2021estimating}}}
         \\
        \max\left[
        \log_2\left(\frac{1}{\left(\eta-1\right)N}\right)-1/\ln{2}+2h\left(\sqrt{1+\left(\eta-1\right)^2N^2}\right),0
        \right]
        & \mbox{if $\eta>1$};
        \\
         \max\left[
        \log_2\left(\frac{1}{N_{\rm add}}\right)-1/\ln{2}+2h\left(\sqrt{1+N_{\rm add}^2}\right),0
        \right]
        & \mbox{if $\eta=1, (1-\eta)N\to N_{\rm add}$}
        \end{array} \right.
\end{align}
\end{widetext}
Here we have defined 
\be 
\nonumber
h(x) =\left(\frac{x+1}{2}\right)\log_2(\frac{x+1}{2})-\left(\frac{x-1}{2}\right)\log_2(\frac{x-1}{2}). 
\ee 

For the teleportation case, we will consider the same values of $\kappa$ that maximizes the quantum capacity lower bound
\be 
\kappa^\star=\arg\max_{\kappa} Q_{\rm LB}(\kappa^2,N_{\rm TP})
\ee 
when we evaluate the corresponding quantum capacity upper bound
\be 
Q_{\rm TP}^{\rm (UB)}=\min[Q_{\rm PLOB}(\kappa^{\star2},N_{\rm TP}), Q_{\rm DE}(\kappa^{\star2},N_{\rm TP}) ].
\ee

\begin{figure}[t]
\centering
\includegraphics[width=0.45\textwidth]{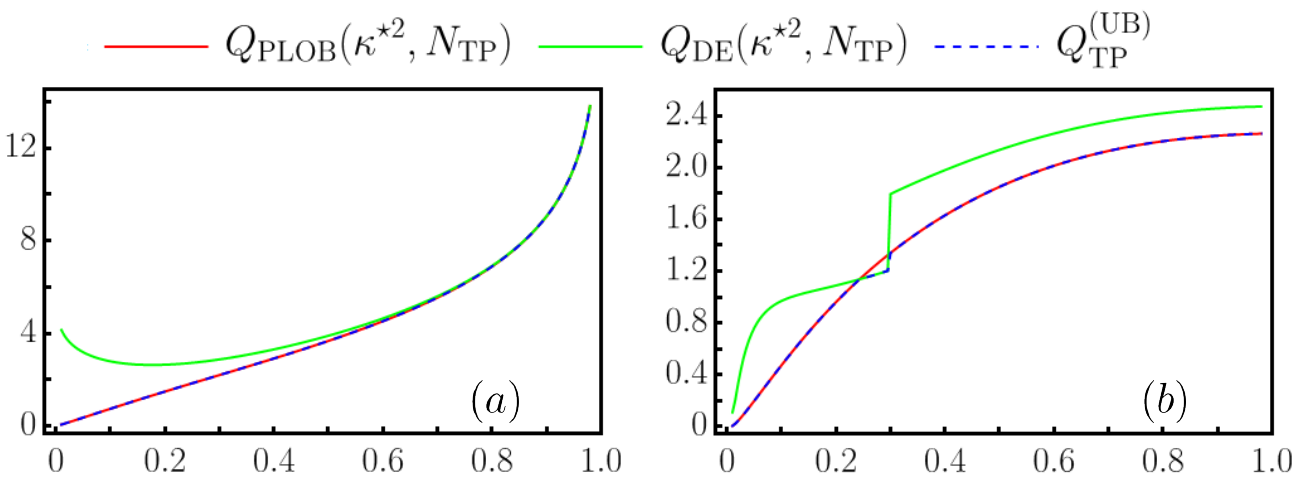}
\caption{Different upper bounds for (a)$\zeta_{\rm{o}}=1$, $\zeta_{\rm{m}}=1$, and (b) $\zeta_{\rm{o}}=0.9$, $\zeta_{\rm{m}}=0.95$, and $n_{\rm{in}}=0.2$.
\label{fig:Upper-bounds}
}
\end{figure}

\begin{figure}[t]
\centering
\includegraphics[width=0.45\textwidth]{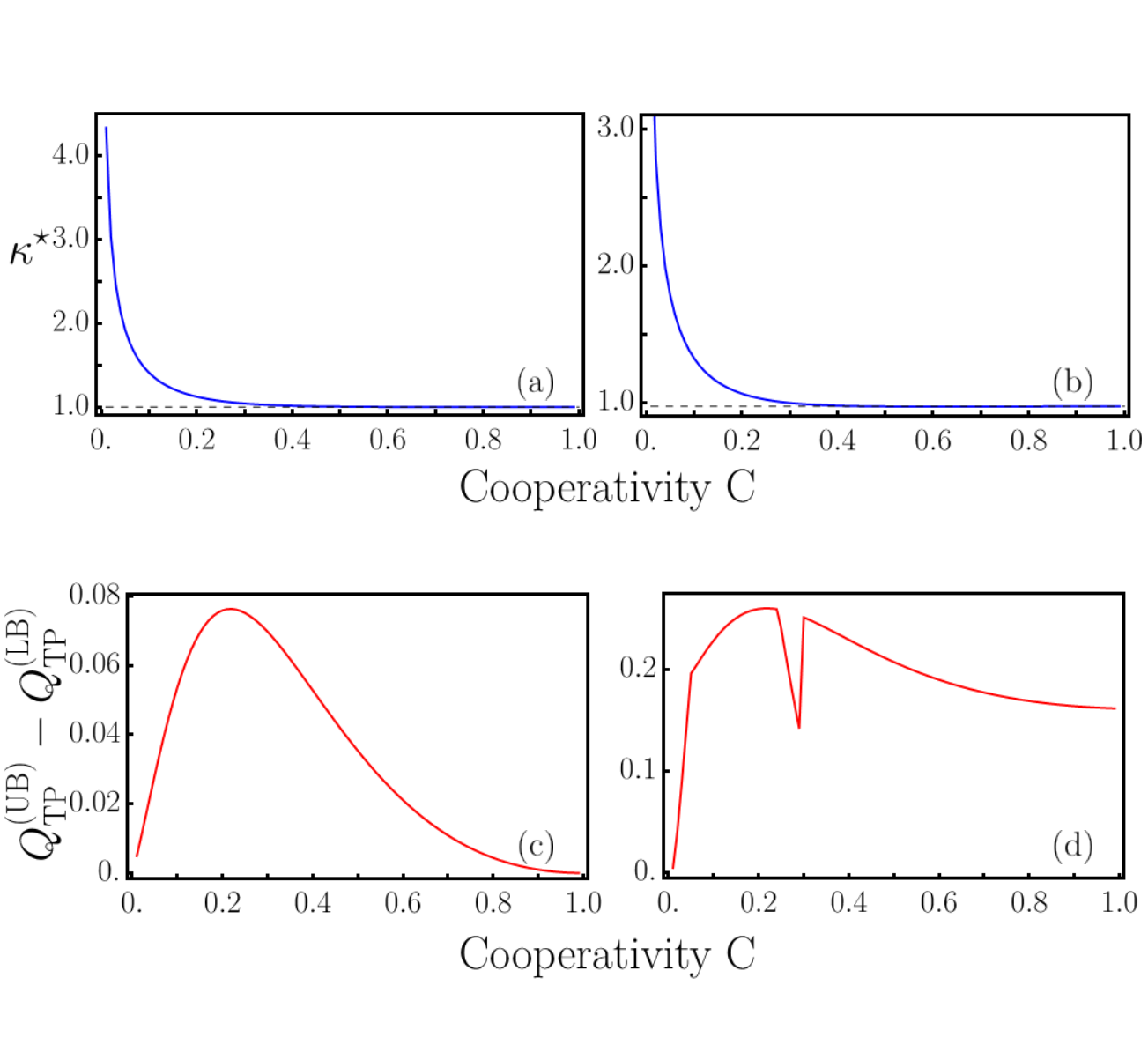}
\caption{Optimum $\kappa$ and difference between upper bound and lower bound. The black dashed lines are \JW{$\kappa=\sqrt{\zeta_{\rm o}/\zeta_{\rm m}}$.} (a)(c) $\zeta_{\rm m}=1$ and $\zeta_{\rm o}=1$. (b)(d) $\zeta_{\rm m}=0.95$, $\zeta_{\rm o}=0.9$ and \JW{$n_{\rm in}=0.2$.}
\label{fig:OptK-Qdiff}
}
\end{figure}

The capacity upper and lower bounds with $\kappa^\star$ are presented in Fig.~\ref{fig:capacity} of the main paper. \JW{We also present different upper bounds in Fig.~\ref{fig:Upper-bounds}.} In Fig.~\ref{fig:OptK-Qdiff}, we present additional data for Fig.~\ref{fig:OptK-Qdiff}(a)(b): the optimum $\kappa^\star$ that maximizes the lower bound and the difference $Q_{\rm TP}^{\rm (UB)}-Q_{\rm TP}^{\rm (LB)}$ between the upper bound and lower bound for the teleportation scheme. Indeed, as discussed in Appendix~\ref{app:channel_definition}, at the limit of $C \rightarrow 1$, we need \JW{$\kappa=\sqrt{\zeta_{\rm o}/\zeta_{\rm m}}$} so that $N_{\rm TP}$ does not diverge. 


\section{Fidelity evaluations} 
\label{app:fidelity}

For a pure input state $\hat{\rho}^{\rm in}$, we can also obtain the fidelity $\mathcal{F}\left(\hat{\rho}^{\rm in},\hat{\rho}^{\rm out}\right)={\rm tr} \left(\hat{\rho}^{\rm in} \hat{\rho}^{\rm out}\right)$ between the input $\hat{\rho}^{\rm in}$ and the output state $\hat{\rho}^{\rm out}$ from their Wigner function
\begin{align}
    \mathcal{F}
    & = 2\pi \int \diff{\bx_B}\; W^{\rm {out}}(\bx_B)W^{\rm {in}}(\bx_B).
\end{align}

For direct conversion, the fidelity can be evaluated from the Wigner function relation in Eq.~\eqref{wigner;thermal_attenuator}, with parameters in Eqs.~\eqref{eta_DC} and~\eqref{N_DC} of the main paper. For teleportation based transduction, the fidelity can be evaluated from the Wigner function relation in Eq.~\eqref{eq:teleported-average-state}. Note that the average fidelity is equal to the fidelity of the average state.

\subsection{Coherent state}
A coherent state $\ket{\alpha}$ is determined by a complex number $\alpha = \alpha_R + i \alpha_I$. Let us use the notation $\balpha = \sqrt{2}(\alpha_R , \alpha_I)^{\rm T}$, then its Wigner function is
\begin{equation}
    W^{\rm {in}}(\bx;\ket{\alpha}) = \frac{1}{\pi} e^{-(\bx-\sqrt{2}\balpha)^2}.
    \label{eq:Fidelity-Wignerfunction}
\end{equation}

In the case of direct conversion (a thermal-attenuator characterized by $\eta_{\rm DC}$ and $N_{\rm DC}$), we get
\begin{equation}
    \mathcal{F} = \frac{1}{1+N_{\rm DC}(1-\eta_{\rm DC})}\exp{-\frac{2\balpha^2(1-\sqrt{\eta_{\rm DC}})^2}{1+N_{\rm DC}(1-\eta_{\rm DC})}}.
\end{equation}
The average fidelity between input and output states in the case of teleportation is given by:
\begin{equation}
    \mathcal{F} = \frac{2}{A\left({\kappa,u,v,w}\right)}\exp{-\frac{2\balpha^2(\kappa-1)^2}{A\left({\kappa,u,v,w}\right)}},
\end{equation}
where we define $A\left({\kappa,u,v,w}\right)\equiv(u+1)\kappa^2-2v\kappa+w+1$.

\subsection{Cat state}

The cat states are defined as $\ket{cat_{\pm}}\equiv N_{\pm}(\ket{\alpha}\pm\ket{-\alpha})$, where $N_{\pm}=(2\pm 2 e^{-2\balpha^2})^{-1/2}$ are normalization constants. The corresponding Wigner functions and the fidelity between the input and output are obtained similarly:
\begin{widetext}
\begin{align}
    &W^{\rm {in}}(\bx;\ket{cat_{\pm}})=N_{\pm}^2\frac{1}{\pi}\left[e^{-(\bx-\sqrt{2}\balpha)^2}+e^{-(\bx+\sqrt{2}\balpha)^2} \pm 2 e^{-\bx^2}\cos[2\sqrt{2}(-q\alpha_I+p\alpha_R)]\right],\\
% \begin{align}
%     &\mathcal{F} = \frac{4N^4_{\pm}}{1+a+b^2}\left(e^{-\frac{2\balpha^2(1-b)^2}{1+a+b^2}}+e^{-\frac{2\balpha^2(1+b)^2}{1+a+b^2}} \pm 2e^{-\frac{2\balpha^2(2+a)}{1+a+b^2}}
%     \right.
%     \nonumber
%     \\
%     &\left.
%     \pm 2e^{-\frac{2\balpha^2(a+2b^2)}{1+a+b^2}}+e^{-\frac{2\balpha^2[2a+(1+b)^2]}{1+a+b^2}}+e^{-\frac{2\balpha^2[2a+(1-b)^2]}{1+a+b^2}}\right)
% \end{align}
    &\mathcal{F} = \frac{4N^4_{\pm}}{1+a+b^2}\left(e^{-\frac{2\balpha^2(1-b)^2}{1+a+b^2}}+e^{-\frac{2\balpha^2(1+b)^2}{1+a+b^2}} \pm 2e^{-\frac{2\balpha^2(2+a)}{1+a+b^2}}
    \pm 2e^{-\frac{2\balpha^2(a+2b^2)}{1+a+b^2}}+e^{-\frac{2\balpha^2[2a+(1+b)^2]}{1+a+b^2}}+e^{-\frac{2\balpha^2[2a+(1-b)^2]}{1+a+b^2}}\right),
\end{align}
\end{widetext}
where $a=(1+2N_{\rm DC})(1-\eta_{\rm DC})$ and $b=\sqrt{\eta}$ for direct conversion; $a = u\kappa^2-2v\kappa+w$ and $b=\kappa$ for teleportation.

\begin{figure}[t]
    \centering
    \includegraphics[width=0.45\textwidth]{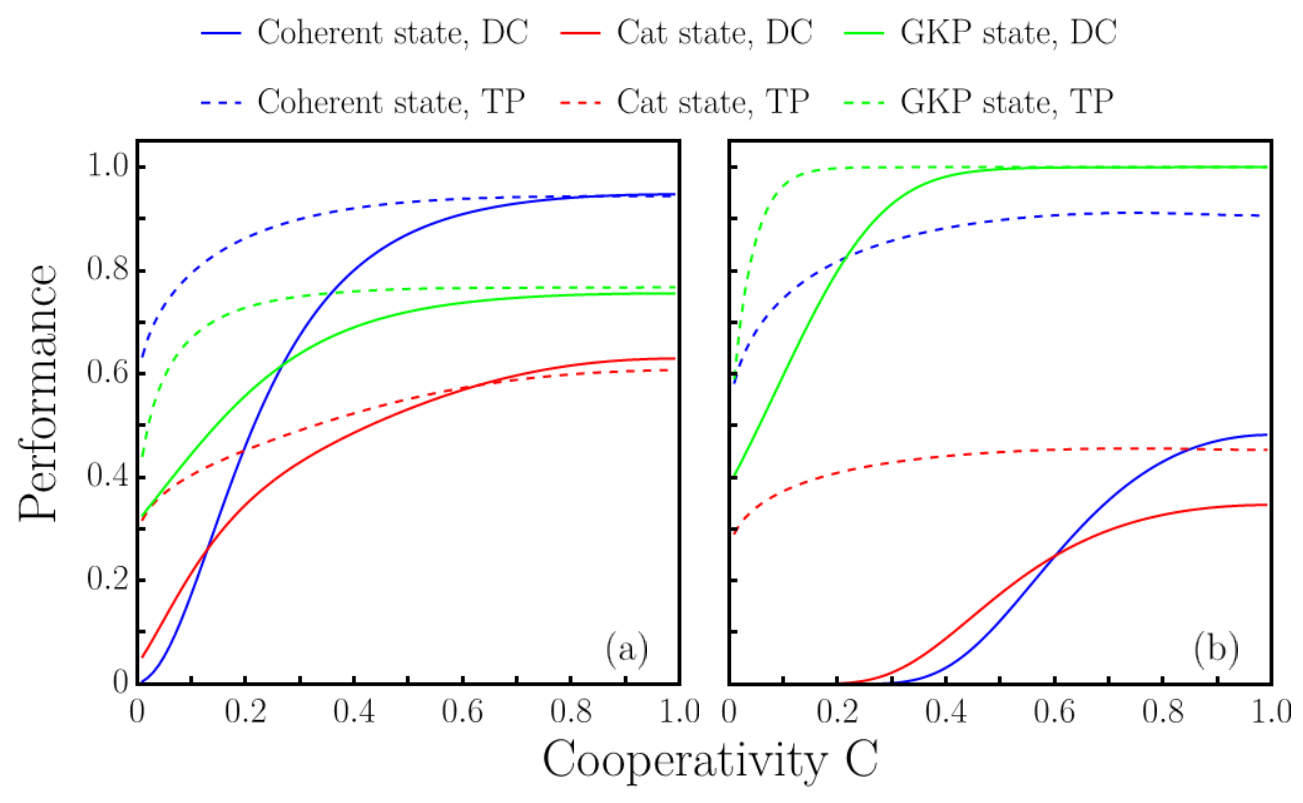}
   \caption{Fidelity for coherent state $\ket{\alpha}$, cat state $N_{+}(\ket{\alpha}+\ket{-\alpha})$ and success probability of GKP state transduction, with finite squeezing noise $\sigma_{\text{GKP}}$. We choose $\zeta_{\rm m}=0.95$, $\zeta_{\rm o}=0.9$ and \JW{$n_{\rm in}=0.2$.} (a) $\alpha=2$ and $\sigma_{\rm{GKP}}=0.4$ (2.2 dB of squeezing) (b)$\alpha=8$ and $\sigma_{\rm{GKP}}=0.1$ (17.0 dB of squeezing). Dashed lines are for teleportation (TP) based transduction and solid lines are for direct conversion (DC).
   }
   \label{fig:fidelity_performance_alpha_change}
\end{figure}

In the main paper, we conclude that when $\alpha$ is large, the crossover of performance in Fig.~\ref{fig:fidelity_performance}(b) will not happen. Here we verify it with $\alpha=8$ in Fig.~\ref{fig:fidelity_performance_alpha_change}(b). Indeed, we see the teleportation scheme (dashed) is consistently better than the direct transducrion scheme (solid).

\section{Additive noise analyses}
\label{app:additive_noise}
We can utilize the channel concatenation relations in Appendix~\ref{app:concatenation} to convert the transduction channels to additive white Gaussian noise channels.

For the direct conversion case (DC), considering the transmissivity in Eq.~\eqref{eta_DC} and noise Eq.~\eqref{N_DC} of the main paper, we can amplify accordingly before the transduction to obtain the additive noise variance
\begin{align}
    \sigma^2_{\rm DC} & = 1 + \frac{4 {{C}}[n_{\rm in}(1-\zeta_{\rm m})-\zeta_{\rm m}]\zeta_{\rm o}}{(1 + {{C}})^2}.
    \label{sigma_DC}
\end{align}

For the teleportation based transduction approach, we need to consider different values of $\kappa$ to obtain the minimum additive noise. When $\kappa \le 1$, we can amplify prior to transduction to obtain an additive noise variance
\begin{align}
    \sigma^2_{\rm TP} & = \frac{1}{2}\left[(u-1)\kappa^2-2 v \kappa + 1 + w\right],
    \label{sigma_TP_1}
\end{align}
where $u,v,w$ are defined in Eqs.~\eqref{uvw} of the main paper. Here ``TP'' stands for teleportation. In this case, $\sigma^2_{\rm TP}$ is minimized when $\kappa=\min[1,v/(u-1)]$. Similarly, when $\kappa \ge 1$, we can append a pure-loss channel after transduction to obtain an additive noise variance
\begin{align}
    \sigma^2_{\rm TP} &  = \frac{1}{2}\left[(w - 1)\frac{1}{\kappa^2}-2 v \frac{1}{\kappa} + 1 + u\right],
    \label{sigma_TP_2}
\end{align}
which is minimized when $\kappa=\max[1,(w-1)/v]$.

\begin{figure}[t]
    \centering
    \includegraphics[width=0.45\textwidth]{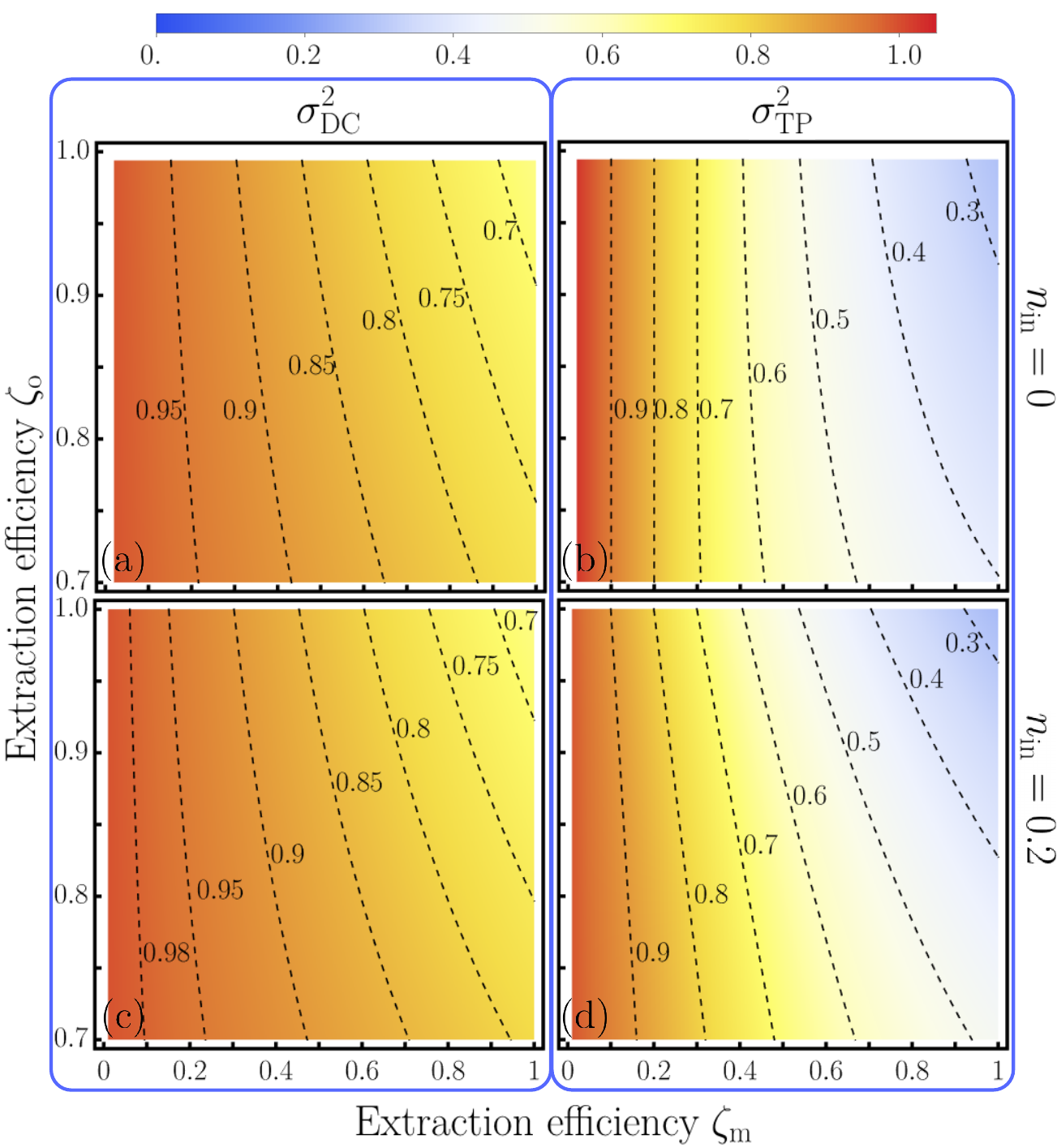}
    \caption{Contour plots of the additive noise variances for cooperativity ${{C}}=0.1$ and (a)(b) $n_{\rm in}=0$ and (c)(d) \JW{$n_{\rm in}=0.2$.}
   \label{fig:additive_noise}
        }
\end{figure}

The comparison between direct conversion and teleportation based schemes can be done by considering Eq.~\eqref{sigma_DC} and Eqs.~\eqref{sigma_TP_1} and~\eqref{sigma_TP_2}. The results are in Fig.~\ref{fig:additive_noise}.

We find that the teleportation based transduction provides a lower additive noise level in the entire parameter region. In particular, at the large cooperativity limit of ${{C}} \to 1$, we have
$
{v}/{(u-1)}={(w-1)}/{v} = \JW{ \sqrt{{\zeta_{\rm o}}/{\zeta_{\rm m}}}. }
$
If $\zeta_{\rm m}>\zeta_{\rm o}$, we take $\kappa = \JW{v/(u-1)}$ and
\begin{equation}
    \sigma^2_{\rm TP} = \frac{1-\zeta_{\rm o}+n_{\rm in}(1-\zeta_{\rm m})}{1+n_{\rm in}(1-\zeta_{\rm m})}.
\end{equation}
While if $\zeta_{\rm m} \le \zeta_{\rm o}$, we take $\kappa =\JW{ (w-1)/v }$ and
\begin{equation}
    \sigma^2_{\rm TP} = \frac{(1+n_{\rm in})(1-\zeta_{\rm m})}{1+n_{\rm in}(1-\zeta_{\rm m})}.
\end{equation}
We can verify that in both cases, $\sigma^2_{\rm DC} \ge \sigma^2_{\rm TP}$ for $C=1$.

\section{GKP error correction} 
\label{app:GKP}

An ideal GKP state is a sum of equal weighted eigenstate of position or momentum quadrature. For the qubit case, up to normalization, we have:
\begin{align}
    &\ket{0}_{\rm GKP}=\sum_{n=-\infty}^\infty \; \ket{\sqrt{\pi}2n}_{\hat{q}},\\
    &\ket{1}_{\rm GKP}=\sum_{n=-\infty}^\infty \; \ket{\sqrt{\pi}(1+2n)}_{\hat{q}}.
\end{align}
A GKP state with finite-squeezing can be modeled as an ideal GKP state with a Gaussian envelope of variance $\sigma_{\rm {GKP}}^2$. As shown in Ref.~\cite{Nohsurface-GKP}, the state is further reduced to an ideal GKP state with additive noise $\sigma_{\rm GKP}$ when $\sigma_{\rm GKP} \ll \sqrt{\pi}$, via random shifts. The level of squeezing equals $1/\Delta$, where
 \begin{equation}
     \frac{1-e^{-\Delta}}{1+e^{-\Delta}} = \sigma_{\rm GKP}^2.
 \end{equation}
 We calculate the squeezing in dB by $10\log_{10}(1/\Delta)$.
 GKP states are resistant to additive noise. For example when there is an displacement error $\hat{q}\rightarrow\hat{q}+\xi_q$, one can measure the displacement of this shifting through a measurement assisted by another GKP state, getting the value of $\xi_q \mod \sqrt{\pi}$, then recover the original state by an anti-displacement. If $\xi_q$ is in the range $[2n\sqrt{\pi}-\sqrt{\pi}/2,2n\sqrt{\pi}+\sqrt{\pi}/2]$ for even $n$, the recovery is perfect. In our case, the displacement $\zeta_{\rm o}$ consists of three independent parts: two from finite GKP states with variances $\sigma_{\rm GKP}^2$ and one from the additive noise channel with variance $\sigma^2$.

Given an additive noise with variance $\sigma^2$, the leading-order success probability for GKP error correction along a single quadrature direction can be obtained as 
\begin{align}
\nonumber
P_{s,p/q} &
= \int_{-\sqrt{\pi}/2}^{\sqrt{\pi}/2}\;\diff{x} \frac{1}{\sqrt{2\pi(\sigma^2+2\sigma_{\text{GKP}}^2)}}\;e^{-\frac{x^2}{2(\sigma^2+2\sigma_{\text{GKP}}^2)}}\\
& = \text{Erf}(\frac{\sqrt{\pi}}{2\sqrt{2(\sigma^2+2\sigma_{\text{GKP}}^2)}}).
\end{align}
The overall success probability is $P_s=P_{s,p/q}^2$.

\end{document}